\documentclass[twocolumn]{aastex7}

\begin{document}

\title{From Giant Envelopes to Rapidly Contracting Cores in Rotating Population III Stars from $500$ to $1000 \, M_\odot$}

\author[orcid=0000-0003-1927-4397,sname='Devesh Nandal']{Devesh Nandal}
\affiliation{Center for Astrophysics $|$ Harvard \& Smithsonian, 60 Garden Street, Cambridge, MA 02138, USA}
\email[show]{deveshnandal@yahoo.com}  
\author[orcid=0000-0002-0302-2577]{John Chisholm}
\affiliation{Department of Astronomy, The University of Texas at Austin, Austin, TX 78712, USA}
\affiliation{Cosmic Frontier Center, The University of Texas at Austin, Austin, TX 78712, USA} 
\email[]{chisholm@austin.utexas.edu}  
\author[orcid=0000-0001-6646-2337]{Daniel J. Whalen}
\affiliation{Institute of Cosmology and Gravitation, Portsmouth University, Dennis Sciama Building, Portsmouth PO1 3FX}
\email{dwhalen1999@gmail.com}  
\author[orcid=0000-0002-2090-9751]{Andreas A.C. Sander}
\affiliation{Institut f{\"u}r Theoretische Physik und Astrophysik, Christian-Albrechts-Universit{\"a}t zu Kiel, Leibnizstr.\ 15, 24118 Kiel, Germany}
\affiliation{Zentrum für Astronomie der Universität Heidelberg, Astronomisches Rechen-Institut, 
   Mönchhofstr. 12-14, 69120 Heidelberg, Germany}
\email[]{asander@astrophysik.uni-kiel.de}  
\author[orcid=0000-0003-0212-2979,gname=Volker,sname=Bromm]{Volker Bromm}
\affiliation{Department of Astronomy, The University of Texas at Austin, Austin, TX 78712, USA}
\email[]{vbromm@astro.as.utexas.edu}
\affiliation{Cosmic Frontier Center, The University of Texas at Austin, Austin, TX 78712, USA} 
\affiliation{Weinberg Institute for Theoretical Physics, University of Texas at Austin, Austin, TX 78712, USA}
\author[0000-0002-1333-147X]{Peixin Zhu}
\affiliation{Center for Astrophysics $|$ Harvard \& Smithsonian, 60 Garden Street, Cambridge, MA 02138, USA}
\email[]{peixin.zhu@cfa.harvard.edu}
\author[orcid=0000-0001-8152-3943]{Lisa J. Kewley}
\affiliation{Center for Astrophysics $|$ Harvard \& Smithsonian, 60 Garden Street, Cambridge, MA 02138, USA}
\affiliation{Research School for Astronomy and Astrophysics, Australian National University, Cotter Road, Weston Creek, ACT 2611, Australia}
\email{lisa.kewley@cfa.harvard.edu}
\begin{abstract}

Population III (Pop III) stars link the chemical enrichment of the early Universe to the formation of the first massive black holes, yet their evolution remains poorly understood at extreme masses. Using GENEC, we evolve rotating zero metallicity models from $500$ to $1000\,M_\odot$ at $v/v_{\rm crit}=0.1$ through core Si burning, and a $1000\,M_\odot$, $v/v_{\rm crit}=0.4$ model evolved through He exhaustion. In the fiducial $1000\,M_\odot$ model, convection and rotational transport carry freshly synthesized C and O into H burning layers, producing $62.1\,M_\odot$ of primary N by He exhaustion. Surface mass loss removes $20.2\,M_\odot$, including $1.8\times10^{-3}\,M_\odot$ of N, with $\log(\mathrm{N}/\mathrm{O})=0.68$, $\log(\mathrm{N}/\mathrm{C})=1.48$, and $\log(\mathrm{C}/\mathrm{O})=-0.80$. By the final hydrostatic model, the inner $100\,M_\odot$ is confined within $0.088\,R_\odot$ while the photosphere remains several thousand solar radii away. As C, Ne, O, and Si burning accelerate, chemical mixing freezes out and advanced products remain buried beneath the envelope. At core Si exhaustion, the central angular velocity reaches $4\times10^{-3}\,{\rm s^{-1}}$ while the surface rotates at $5.25\times10^{-8}\,{\rm s^{-1}}$, a contrast of $7.6\times10^4$. Contraction drives this spin up, while inner specific angular momentum changes by less than $1\%$. The maximum $j_{\rm eq}/j_{\rm ISCO}$ is $0.313$, limiting disk formation in our inside out estimate. These models show that Pop~III stars can produce strong primary N signatures and rapidly rotating black hole progenitors while their envelopes determine which products enrich the early Universe.

\end{abstract}


\section{Introduction}
\label{sec:introduction}

Pop III stars formed from metal-free gas under physical conditions that differed fundamentally from those of later stellar generations \citep{Bromm2004,Yoshida2006}. In the absence of metal and dust cooling, the thermal evolution of primordial gas is controlled primarily by molecular hydrogen chemistry and cooling, and densities and radiation backgrounds regulate how efficiently the gas can cool and fragment \citep{Machacek_2001,Schauer_2021}. Numerical simulations show that primordial clouds can fragment into multiple accreting protostars rather than a single one, although the resulting top-heavy initial mass function (IMF) can still be described with a characteristic stellar mass \citep{Greif2011,Greif2012,latif22a}. The subsequent accretion history and radiative feedback then determine which objects continue to grow \citep{Stacy2012,Hirano2014,Hosokawa2016}. Under conditions that sustain sufficiently large inflow rates, ionizing UV from the star can permit continued accretion for up to several hundred solar masses and beyond \citep{Woods2017,Nandal2023,Nandal2024d,Gurian2026}. Such stars occupy the upper end of the very massive star (VMS) regime and the lower end of what we refer to here as the extremely massive star (EMS) regime. Their evolution is relevant not only to the formation of the first massive black holes, but also to the radiative and chemical feedback that shaped the first galaxies \citep{wet08b,wet08a,jet09b,wf12,smidt18,Regan_2020b,Volonteri2021,latif21a,latif22b,latif26a,wet26a}. No single Pop III star has yet been identified observationally, making theoretical predictions of their structures, chemical signatures, and final evolution particularly important for connecting models of the first stars with observations \citep{Bromm2013,KlessenGlover2023}.

The internal evolution of these VMSs enters a different regime from that of ordinary massive stars because of their large luminosities and strong radiation pressure support \citep{Vink2018}. Their interiors contain exceptionally large nuclear burning regions and approach conditions where relatively small changes in the equation of state can have important structural consequences. At sufficiently high temperatures, electron positron pair creation removes thermal energy from the radiation field and softens the equation of state, which can drive contraction and in some cases explosive nuclear burning \citep{Heger2002}. Numerical calculations established the transition from pair instability supernovae toward direct collapse at larger core masses \citep{Heger2003} while later studies showed that the boundaries depend sensitively on the stellar structure and nuclear physics \citep{Takahashi2018,Farmer2019}. \citet{Ohkubo2006} followed nonrotating $500$ and $1000\ \, M_\odot$ Pop III stars through advanced burning toward iron core collapse. More recent calculations have considered primordial stars up to $1000\ \, M_\odot$ with updated stellar physics, although many terminate once a global pair softening criterion is reached \citep{volp23}. The hydrostatic evolution connecting He exhaustion to the final advanced core therefore remains comparatively poorly explored across this extreme range in mass
\citep{Ohkubo2009,Yoon2012}.

Rotation introduces a second set of coupled effects because it changes both chemical transport and the redistribution of angular momentum \citep{Zahn1992,Maeder1997}. Meridional circulation and shear transport can move fresh fuel and nuclear products across radiative regions, allowing rotational mixing to alter burning lifetimes, the extent of chemically processed regions, and the surface composition. At the same time, angular momentum is redistributed between the contracting interior and the expanding envelope, which determines how much specific angular momentum survives in the core \citep{Heger2000,Meynet2015}. These effects are particularly important at very low metallicity because weak radiative winds reduce both mass and angular momentum loss from the stellar surface \citep{Kudritzki2002,Ekstrom2008}. Intriguingly, cosmological simulations of Pop~III star formation, tracking the angular momentum content of the growing protostars, have predicted surface rotation velocities close to the critical value \citep{Stacy_rot2011,Stacy_rot2013}.  

A star that loses less angular momentum can remain rotationally mixed for a larger fraction of its nuclear evolution, whereas stronger winds can remove high angular momentum surface material and spin the star down. Calculations of rotating primordial VMSs have demonstrated substantial changes in chemical evolution and retained angular momentum \citep{Yoon2012}, while finite metallicity models illustrate how strongly the evolution changes once wind mass loss becomes efficient \citep{Yusof2013}. The final structure of an extremely massive primordial star therefore depends not only on its initial mass, but also on how rotation and mass loss couple the long lived burning phases to the rapidly contracting core \citep{Meynet2016,Sibony2024,Nandal2024b}.

This connection becomes particularly important after core He exhaustion. In classical massive stars, the carbon abundance left by He burning strongly influences the subsequent C burning history and the shell structure that develops before collapse \citep{Sukhbold2014,Patton2020}. Changes in the residual C abundance alter where and how efficiently C burns, which modifies the density and composition profiles inherited by the later burning stages. The resulting presupernova structure can therefore vary nonmonotonically with initial mass even when neighboring stars begin with similar properties \citep{Sukhbold2018,Sukhbold2020}. Rotating low metallicity calculations further show that these structural changes alter the final elemental yields and the masses and compositions of the cores that enter collapse \citep{Limongi2018}. Following the evolution through Si burning is particularly important because the nuclear timescales contract from thousands of years after He exhaustion to days, hours, and eventually minutes. Neutrino cooling and rapid contraction then increasingly limit the time available for chemical and angular momentum transport, allowing differences inherited from the earlier evolution to become frozen into the terminal structure. Recent extensions of GENEC provide the thermodynamic treatment and advanced nuclear network required to follow this sequence consistently through Si burning \citep{Griffiths2025}. Whether the structural sensitivity familiar from lower mass stars persists or becomes amplified at several hundred solar masses has not yet been systematically established for rotating primordial stars \citep{Yoon2012,volp23,Sibony2024}.

The chemical consequences are equally important because rotational transport can create primary nitrogen even when the star begins without CNO nuclei \citep{Ekstrom2008,Tsiatsiou2024}. Very massive and supermassive Pop III models can produce particularly large nitrogen reservoirs \citep{Nandal2024,Nandal2025}, making them relevant to the unusual abundance patterns recently discovered in several high redshift systems \citep{Bunker2023,Cameron2023,Senchyna2023,MarquesChaves2024}. These abundance measurements are themselves sensitive to the physical conditions of the emitting gas. Ionization structure and electron density affect the nebular diagnostics used to infer elemental abundances, and recent calculations show that inappropriate density assumptions can systematically bias the inferred O/H and N/O ratios in compact high redshift systems \citep{Kewley2013,Martinez2025}. The comparison between stellar models and observed abundance patterns therefore requires both the stellar production of the elements and a consistent interpretation of the nebular gas.

Producing an element inside a star is also not equivalent to ejecting it. Newly synthesized material must first reach layers that can be removed, and a physical mass loss or stripping mechanism must then carry that material away from the star \citep{Vink2018,Nandal2024e,Nandal2026b}. This distinction becomes increasingly important during advanced burning because the newly synthesized Si and Fe group material can remain confined to a compact interior beneath a massive H and He rich envelope. Such material can dominate the terminal hydrostatic composition of the core while contributing little to the surrounding environment if the overlying layers are not removed before or during collapse. The chemical contribution of these stars therefore depends on the depth reached by mass removal as well as on the nucleosynthesis itself.

The angular momentum of the compact interior provides a complementary connection to the final fate. Rapidly rotating VMSs have been considered as possible progenitors of black holes surrounded by accretion disks when the collapsing material retains sufficient specific angular momentum \citep{Fryer2001}. Relativistic collapse calculations show that material with sufficiently large specific angular momentum can circularize around the growing black hole and form a massive disk capable of driving associated outflows \citep{Uchida2019,Shibata2025}. Rotating Pop III calculations have also identified possible disk forming progenitors at lower masses \citep{Volpato2024}. A large central angular velocity alone is not sufficient, however, because contraction can produce rapid spin without increasing the specific angular momentum of the collapsing material. The relevant quantity is therefore the retained angular momentum profile and whether individual shells contain enough specific angular momentum to circularize around the growing black hole \citep{Yoon2005,Woosley2006}. Rotation can consequently affect the final fate only if sufficient angular momentum survives the preceding transport and mass loss history.

In this work we use the extended GENEC framework to evolve rotating Pop III stars with initial masses from $500$ to $1000\ \, M_\odot$ through the advanced hydrostatic burning stages \citep{Eggenberger2008,Griffiths2025,Nandal2025b}. We determine how the long H and He burning phases establish the chemical and rotational structure inherited by the advanced evolution, and how the rapidly shortening C, Ne, O, and Si burning stages subsequently reshape the compact interior. We follow the global and central evolution in Section~\ref{sec:global_central_evolution}, before examining the chemical and structural assembly of the advanced interior and its terminal structure in Sections~\ref{sec:advanced_interior} and \ref{sec_terminal_structure}. Sections~\ref{sec_rotation}--\ref{sec:deeper_stripping} then examine the angular momentum evolution, the global elemental inventory, and the distinction between material that is synthesized and material that can actually be removed. Finally, Section~\ref{sec:discussion} considers the consequences for collapse circularization and the freeze-out of chemical mixing during the final burning stages.

\section{Methods}
\label{sec:methods}

We computed the stellar models with the Geneva stellar evolution code GENEC \citep{Eggenberger2008,Nandal2026a}. The grid spans the upper VMS and lower EMS regimes defined above, with Pop III models of initial masses $500$, $600$, $700$, $800$, $900$, and $1000\ \, M_\odot$ at $Z=0$, and an initial primordial composition of $X=0.7516$ and $Y=0.2484$. The standard grid begins with an initial rotation rate of $v/v_{\rm crit}=0.1$. This choice reflects the rotational physics of stars approaching the supermassive regime, whose large Eddington factors allow the $\Omega\Gamma$ limit to be reached already at relatively modest surface rotation \citep{Haemmerle2018,Lionel2021,Nandal2025b}. More rapid initial rotation is possible, but the resulting approach to the $\Omega\Gamma$ limit promotes additional mass and angular momentum loss and subsequent spin down. The adopted value therefore provides a physically motivated reference rotation rate for this mass regime, rather than the $v/v_{\rm crit}\simeq0.4$ commonly used for classical rotating massive star models. We additionally computed a $1000\,M_\odot$ model with $v/v_{\rm crit}=0.4$ through core He exhaustion to provide a fixed mass comparison with stronger initial rotation. Here $v_{\rm crit}$ denotes the critical equatorial velocity, which is reduced near the Eddington limit through the $\Omega\Gamma$ effect \citep{OGlimit2000}.

Rotation is treated with the shellular formalism used in the Geneva models \citep{Zahn1992,Maeder1997,Nandal2025b}. Angular momentum is transported by meridional circulation and shear diffusion, while rotational chemical transport is described by $D_{\rm chem}=D_{\rm shear}+D_{\rm eff}$. The models are nonmagnetic. Convective boundaries are determined with the Schwarzschild criterion and core overshooting of $0.1H_P$ is applied during core H and He burning \citep{Ekstrom2012,Griffiths2025}.

Radiative mass loss follows the prescription of \citet{Vink2001} in the hot stellar regime and \citet{DeJager1988} as the models evolve to cooler temperatures, with the rate evaluated from the instantaneous stellar properties required by each prescription. At $Z=0$, however, classical line driven mass loss is extremely inefficient because the metal line opacity responsible for driving ordinary massive star winds is initially absent \citep{Krt2006,Krt2009}. We retain these radiative prescriptions throughout the evolution, but the resulting rates remain low, typically of order $10^{-6} \, M_\odot \, {\rm yr}^{-1}$. The dominant mass loss in the models is therefore associated with rotation and the approach to the $\Omega\Gamma$ limit \citep{OGlimit2000,Nandal2025b}. When the surface approaches this limit, GENEC removes mass and angular momentum to keep the subsequent model below critical rotation, providing the dominant mechanical mass loss channel in these calculations. Pulsational mass loss is not included because following pulsational growth and discrete ejection events requires a time dependent treatment beyond the hydrostatic framework used here \citep{Baraffe2001,Nandal2026b}.

For the advanced evolutionary stages we use the extended GENEC microphysics in \citet{Griffiths2025}. The equation of state combines the standard GENEC Dichte equation of state with the Timmes electron-positron equation of state at high temperature and density \citep{TimmesSwesty2000}. The latter includes electron-positron pair creation and relativistic electron degeneracy, allowing the thermodynamic derivatives and $\Gamma_1$ to be followed through the pair softened regime. Nuclear burning during H and He burning uses the standard extended GENEC network, while the evolution from core C burning onward uses the 25 species GeValNet25 network. This network follows the advanced alpha chain and includes an effective electron capture sequence through the iron group to approximate the evolution of $Y_e$. GeValNet25 was tested against MESA models evolved with the analogous reduced \texttt{approx21\_plus\_co56} network, producing broadly consistent central evolution and core properties while retaining some differences in the detailed shell structure \citep{Griffiths2025}. Its detailed partition within the iron group is therefore not interpreted as precision isotopic nucleosynthesis.

GENEC uses an adaptive Lagrangian mass mesh whose number of shells changes during the evolution. Shells are introduced where variations in the stellar structure require greater spatial resolution and are merged where the structure becomes sufficiently smooth \citep{Griffiths2025}. The present calculations contain approximately $2000$ mass shells on average, compared with roughly $1500$ shells typical of recent advanced GENEC calculations. The increased resolution helps resolve the narrow burning regions and steep composition gradients that develop during advanced burning. The shell number should therefore be interpreted as adaptive structural resolution rather than as a uniform mass spacing.

For consistency between models, central H, He, C, Ne, and O exhaustion are defined to be when the corresponding active central fuel abundance falls below $10^{-4}$. Central Si exhaustion is defined using the advanced $^{28}{\rm Si}$ abundance at the first post peak crossing below $10^{-3}$. The $800\,M_\odot$ calculation does not reach this final criterion. GENEC remains hydrostatic throughout these calculations, so the terminal structures represent the latest computed hydrostatic progenitors rather than models of the subsequent dynamical collapse.
\begin{figure*}[!t]
\centering
\begin{minipage}[c]{0.49\textwidth}
    \centering
    \includegraphics[width=\linewidth]{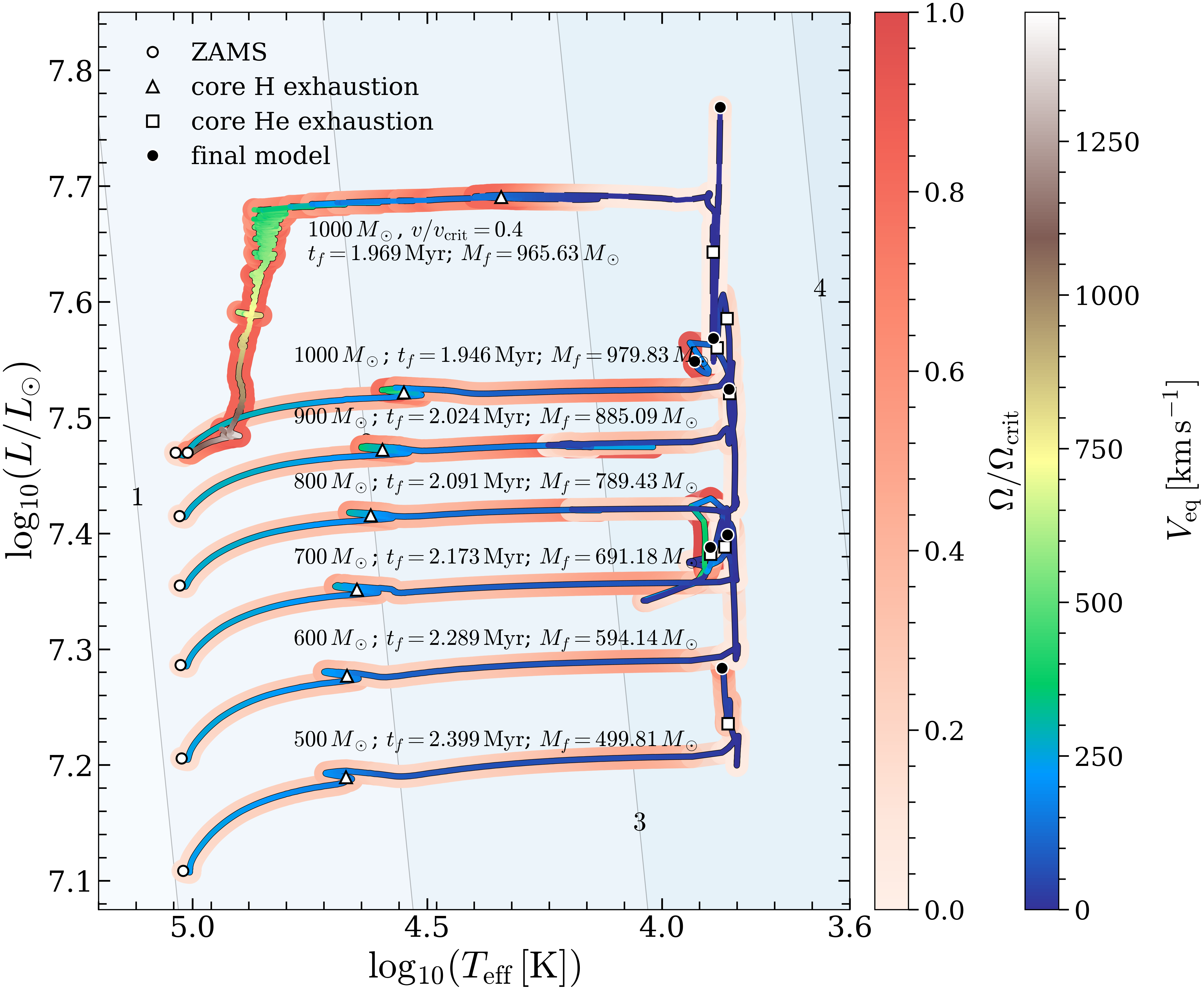}
\end{minipage}
\hfill
\begin{minipage}[c]{0.49\textwidth}
    \centering
    \includegraphics[width=\linewidth]
    {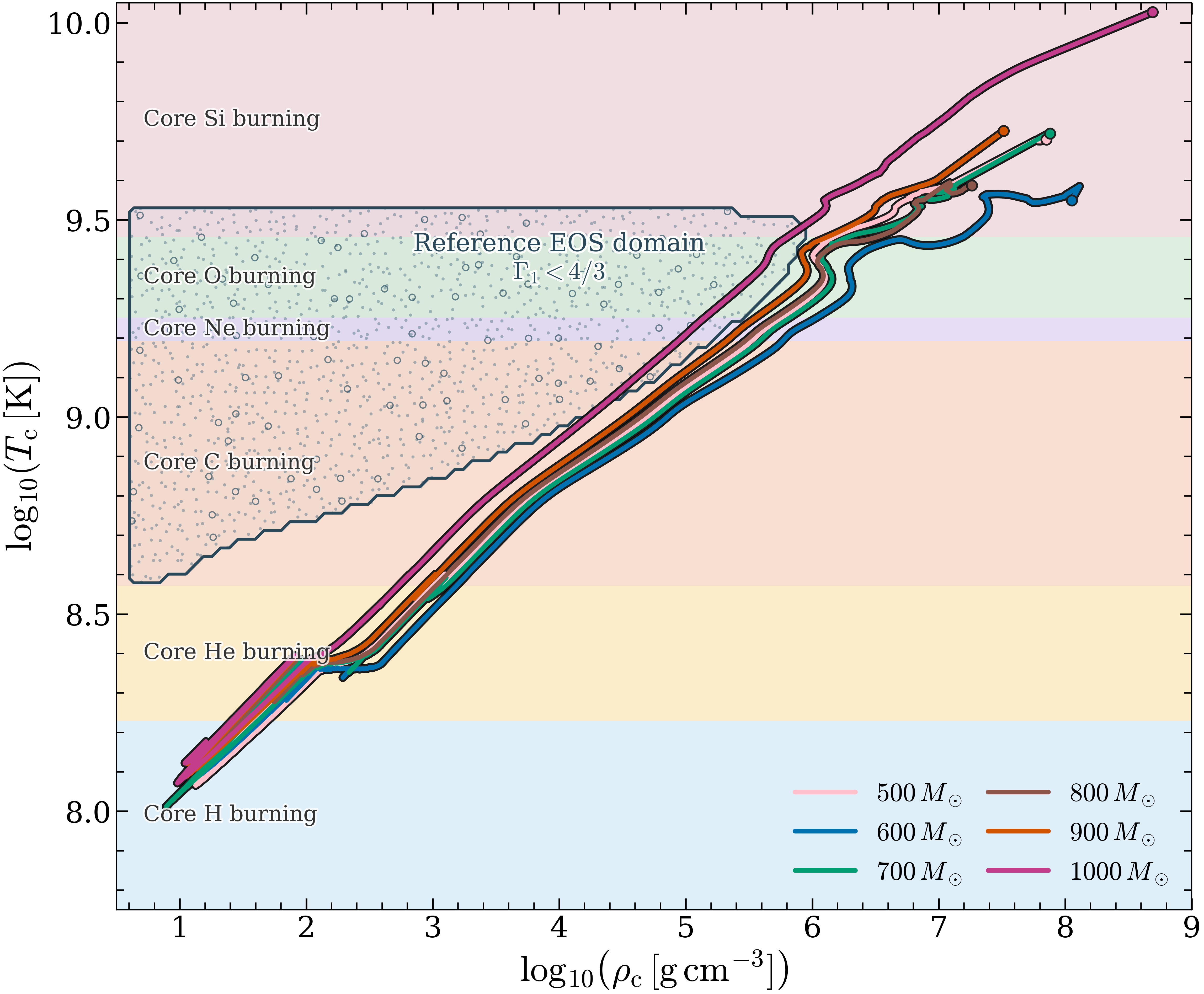}
\end{minipage}

\caption{
Global and central evolution of the rotating Pop III models.
The left panel shows $\log(L/L_\odot)$ against $\log(T_{\rm eff}/{\rm K})$ for the six models with initial masses from $500$ to $1000\,M_\odot$ and $v/v_{\rm crit}=0.1$, together with the $1000\,M_\odot$ model with $v/v_{\rm crit}=0.4$.
Green background curves mark constant $\log(R/R_\odot)$.
The narrow foreground color gives the surface equatorial velocity $V_{\rm eq}$ and the broad halo gives $\Omega/\Omega_{\rm crit}$ according to the two color bars.
Open circles mark the ZAMS, triangles mark central H exhaustion, squares mark central He exhaustion, and filled circles mark the final model.
Track labels give the initial mass, final age, and final mass.
The right panel shows $\log(T_{\rm c}/{\rm K})$ against $\log(\rho_{\rm c}/{\rm g\,cm^{-3}})$ for the six models with $v/v_{\rm crit}=0.1$.
Track colors identify the initial masses.
The horizontal shaded regions mark the labelled H, He, C, Ne, O, and Si burning temperature intervals.
The irregular stippled region marks the reference equation of state domain with $\Gamma_1<4/3$, where the adiabatic pressure response is sufficiently softened that a radiation pressure dominated self-gravitating region can become dynamically unstable.
Filled circles mark the final point of each central trajectory. The more luminous $1000\,M_\odot$ track shows the $v/v_{\rm crit}=0.4$ comparison model, which is evolved through core He exhaustion. 
}
\label{fig:global_central_evolution}
\end{figure*}

\section{Results}
\label{sec:results}

We first follow how initial mass and rotation shape the stellar evolution from core H burning through the final advanced burning stages. We then examine how this history determines the internal structure, angular momentum distribution, and chemical inventory of the final hydrostatic models.

\subsection{Nuclear burning evolution}
\label{sec:global_central_evolution}

Figure~\ref{fig:global_central_evolution} combines the evolution in the Hertzsprung Russell (HR) diagram with the central trajectories in the $\rho_{\rm c}$ and $T_{\rm c}$ plane. Table~\ref{tab:evolution_stages} gives the corresponding nuclear timescales, maximum central convective core masses, and surface equatorial velocities. We will now explore the evolution of models through each evolutionary stage.

\paragraph{Core hydrogen burning} All models begin their evolution from the zero age main sequence (ZAMS) on the left side of the HR diagram. The ZAMS luminosity increases from $\log(L/L_\odot)=7.109$ at $500\,M_\odot$ to $7.470$ at $1000\,M_\odot$, while $T_{\rm eff}$ changes only from $1.05\times10^{5}$~K to $1.09\times10^{5}$~K. The corresponding radius increases from $10.9$ to $15.3\,R_\odot$ and $V_{\rm eq}$ rises from $242$ to $286~{\rm km\,s^{-1}}$. The ZAMS effective temperature therefore changes by only about $4$\% across the sequence, so the increase in luminosity is primarily due to the larger stellar radius \citep[e.g.,][]{BKL2001}.

The redward evolution begins progressively earlier in the nuclear evolution as the initial mass increases. We quantify this using the first crossing of $\log(T_{\rm eff}/{\rm K})=4.6$. The $800$, $900$, and $1000\,M_\odot$ models reach this point while their central H mass fractions are still $0.109$, $0.134$, and $0.146$, respectively, whereas the $500$, $600$, and $700\,M_\odot$ models cross only after central H exhaustion. Thus the three highest mass models begin the strong envelope expansion before the end of H burning, while the lower mass models remain hotter until the beginning of He burning. A similar mass dependence is found in nonrotating VMS calculations, where more massive stars contain larger nuclear burning regions and undergo stronger envelope expansion \citep{Nandal2023}.  

The central convective cores are already exceptionally large during this phase. Their maximum masses increase from $466\,M_\odot$ at $500\,M_\odot$ to $950\,M_\odot$ at $1000\,M_\odot$, corresponding to about $93$\% to $95$\% of the initial stellar mass. Rotation therefore acts on stars whose nuclear evolution is dominated by a very large mixed central region. Meridional circulation advects angular momentum during the main sequence while shear and chemical transport operate diffusively, so rotational mixing can connect regions outside the formal convective boundary to the central burning region.

The fixed mass rotation comparison shows this directly. The more rapidly rotating ($v/v_{\rm crit}=0.4$) $1000\,M_\odot$ model crosses $\log T_{\rm eff}=4.6$ only when $X_{\rm c}({\rm H})=0.043$, compared with $0.146$ in the reference ($v/v_{\rm crit}=0.1$) model. Its H burning lifetime is also longer at $1.7505$~Myr instead of $1.7010$~Myr. The more rapidly rotating model therefore remains hot until substantially more central H has been consumed.

The surface composition provides an independent signature of this mixing. At H exhaustion the reference $1000\,M_\odot$ model retains a surface H mass fraction of $0.751$, whereas the $v/v_{\rm crit}=0.4$ model has already fallen to $0.273$. The corresponding maximum H burning convective core masses remain similar at $950.2\,M_\odot$ and $924.8\,M_\odot$, so the much stronger surface H depletion in the more rapidly rotating model cannot be explained by the convective core size alone. The surface rotational evolution also differs strongly between the two models. The $v/v_{\rm crit}=0.4$ model begins with $V_{\rm eq}=1110~{\rm km\,s^{-1}}$ and reaches only $117~{\rm km\,s^{-1}}$ by H exhaustion, while its radius grows from $17$ to $474\,R_\odot$. It has also lost about $34\,M_\odot$ by this stage. The simultaneous expansion and removal of angular momentum rich surface material therefore reduce the absolute surface velocity even though $\Omega/\Omega_{\rm crit}$ remains high at $0.77$. Overall, core H burning is characterized by exceptionally large convective cores and an increasingly early redward evolution as the initial mass increases. At fixed mass, stronger rotation substantially enhances surface H depletion without greatly changing the formal convective core size, while envelope expansion and angular momentum loss drive strong surface spin down.

\begin{figure*}[t]
\centering

\begin{minipage}[c]{0.49\textwidth}
    \centering
    \includegraphics[width=\linewidth]{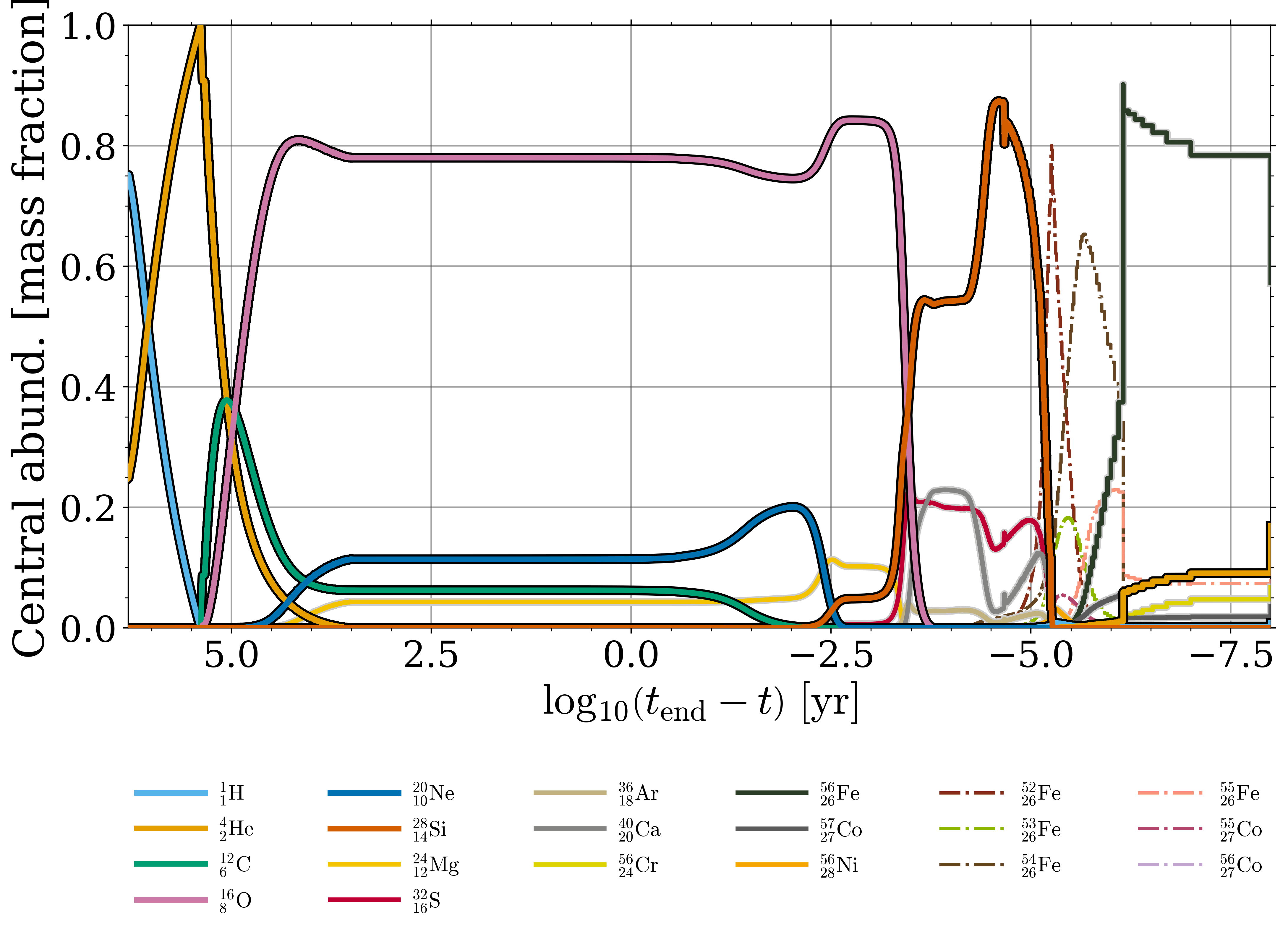}
\end{minipage}
\hfill
\begin{minipage}[c]{0.49\textwidth}
    \centering
    \includegraphics[width=\linewidth]{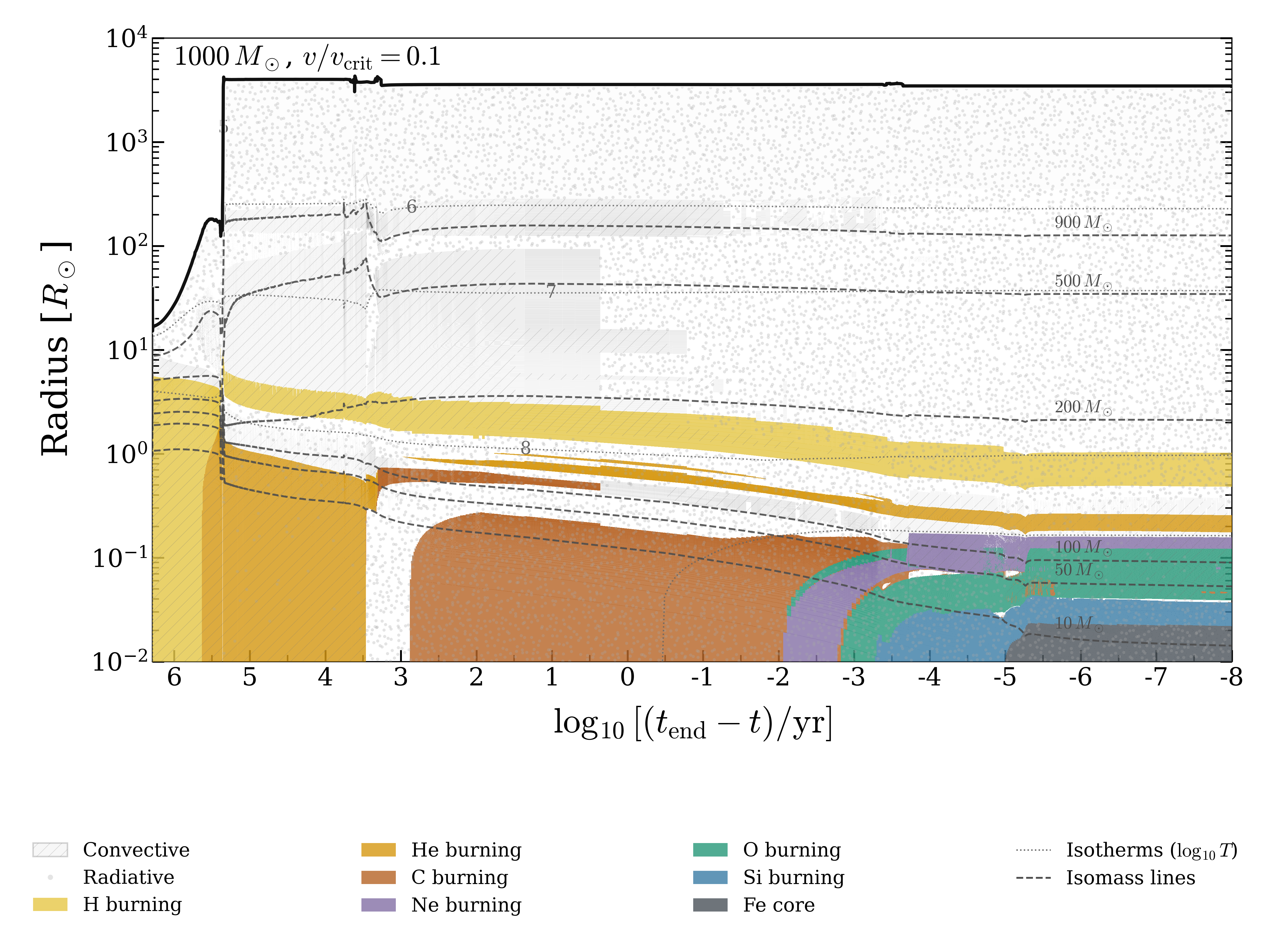}
\end{minipage}

\caption{
Central composition and internal structure of the
$1000\,M_\odot$ model with $v/v_{\rm crit}=0.1$.
Left: central mass fractions for
$^{1}{\rm H}$,
$^{4}{\rm He}$,
$^{12}{\rm C}$,
$^{16}{\rm O}$,
$^{20}{\rm Ne}$,
$^{24}{\rm Mg}$,
$^{28}{\rm Si}$,
$^{32}{\rm S}$,
$^{36}{\rm Ar}$,
$^{40}{\rm Ca}$,
$^{56}{\rm Cr}$,
$^{56}{\rm Fe}$,
$^{57}{\rm Co}$,
and $^{56}{\rm Ni}$ (solid) and the 
additional iron group elements
$^{52}{\rm Fe}$,
$^{53}{\rm Fe}$,
$^{54}{\rm Fe}$,
$^{55}{\rm Fe}$,
$^{55}{\rm Co}$,
and $^{56}{\rm Co}$
(dot-dashed).
Right: stellar radius vs. $\log_{10}[(t_{\rm end}-t)/{\rm yr}]$, where $t_{\rm end}$ is the age of the final hydrostatic GENEC model. Colored regions identify H, He, C, Ne, O, and Si burning together with the Fe group core.
Diagonal hatching marks convective regions while light stippling marks radiative regions.
The dashed curves mark enclosed masses for
$10$, $50$, $100$, $200$, $500$, and $900\,M_\odot$.
The dotted curves mark the indicated isotherms and the solid black upper boundary gives the photospheric radius.
}
\label{fig:advanced_interior}
\end{figure*}

\paragraph{Core helium burning}
The largest change in the surface structure occurs between H and He exhaustion. At He exhaustion the slowly rotating sequence has $T_{\rm eff}\simeq7.2\times10^{3}$~K to $7.8\times10^{3}$~K and radii of about $2.6\times10^{3}\,R_\odot$ to $3.8\times10^{3}\,R_\odot$ (left panel of Figure~\ref{fig:global_central_evolution}). The corresponding $V_{\rm eq}$ values have fallen to only $2$ - $45~{\rm km\,s^{-1}}$.

The decline in surface rotation is primarily connected to the enormous increase in radius. This remains clear even when compared with the total mass removed over the full evolution. The $500\,M_\odot$ model ends at $499.814\,M_\odot$, while the $800\,M_\odot$ model ends at $789.434\,M_\odot$, yet their surface velocities fall from $170$ to $25.5~{\rm km\,s^{-1}}$ and from $239$ to $21.3~{\rm km\,s^{-1}}$ between H and He exhaustion. The strong surface spin down of the standard sequence is therefore driven primarily by envelope expansion rather than by mass removal.

The He burning convective cores remain very large, but their sensitivity to rotation becomes much stronger. Their maximum mass increases from about $250\,M_\odot$ at $500\,M_\odot$ to $477.4\,M_\odot$ at $1000\,M_\odot$ in the standard sequence. For the reference $1000\,M_\odot$ model this is about half of the maximum H burning convective core mass of $950.2\,M_\odot$. The $v/v_{\rm crit}=0.4$ model instead reaches a He burning convective core mass of $779.1 \, M_\odot$, about $84$\% of its H burning value and $63$\% larger than in the reference model. Its surface H mass fraction has also fallen to $0.103$ by He exhaustion, compared with $0.456$ in the reference model. The corresponding surface He, C, N, and O mass fractions are $0.758$, $1.57\times10^{-2}$, $8.07\times10^{-3}$, and $0.104$ in the more rapidly rotating model, compared with $0.518$, $5.70\times10^{-4}$, $2.11\times10^{-2}$, and $4.83\times10^{-3}$ in the reference model. Thus both initially metal free models develop substantial surface CNO enrichment by He exhaustion, with the more rapidly rotating model showing particularly strong C and O enrichment. Despite this more extensive mixing, the He burning interval decreases from $241.45$ to $217.95$~kyr, so stronger rotation does not extend the core He burning lifetime in this comparison.

The surface spin down does not imply that the interior follows the same rotational evolution. At He exhaustion the slow $1000\,M_\odot$ model has $\Omega_{\rm c}=1.17\times10^{-5}~{\rm s^{-1}}$ while its surface rotates at $5.23\times10^{-9}~{\rm s^{-1}}$. The corresponding values in the faster model are $1.11\times10^{-4}$ and $1.70\times10^{-9}~{\rm s^{-1}}$. A large rotational contrast between the contracting interior and the extended envelope is therefore already established before C burning begins. Core He burning also provides sufficient time for chemical transport between the burning regions. Freshly produced C and O are carried outward by the combined action of convection and rotational transport into overlying H burning layers, where they provide the catalysts for CNO cycling and the production of primary N. This coupling is particularly effective during the $\sim10^5$ yr He burning phase and can persist into C burning, before the rapidly shortening advanced burning timescales progressively suppress large scale chemical redistribution.

\paragraph{Core carbon burning}

The evolution accelerates rapidly after He exhaustion. The interval to C exhaustion ranges from $3.16$~kyr - $8.44$~kyr across the standard sequence, while the maximum C burning convective core ranges from about $24\,M_\odot$ - $143\,M_\odot$. The absence of a simple mass ordering already indicates that the advanced core structure cannot be inferred from initial mass alone.

The $1000\,M_\odot$ model illustrates the increasing separation between the core and surface evolution. From He to C exhaustion its central density increases from $\log(\rho_{\rm c}/{\rm g\,cm^{-3}})=2.77$ - $5.12$ and its central temperature rises from $\log(T_{\rm c}/{\rm K})=8.58$ - $9.24$, as shown in the right panel of Figure~\ref{fig:global_central_evolution}. Its radius changes only from about $3760\,R_\odot$ - $3570\,R_\odot$ over the same interval. The star therefore remains in nearly the same cool region of the HR diagram while the central conditions change by several orders of magnitude in density.

Core contraction also begins to dominate the internal rotational evolution. In the $1000\,M_\odot$ model $\Omega_{\rm c}$ rises from $1.17\times10^{-5}~{\rm s^{-1}}$ to $4.32\times10^{-4}~{\rm s^{-1}}$ between He and C exhaustion. The corresponding surface velocity increases only from $13.7~{\rm km\,s^{-1}}$ to $29.2~{\rm km\,s^{-1}}$.

\paragraph{Core neon burning}
The interval from C to Ne exhaustion lasts only $0.282$~d in the $1000\,M_\odot$ model and reaches $10.56$~d in the $600\,M_\odot$ model. The central temperature at Ne exhaustion lies in the narrow range $\log(T_{\rm c}/{\rm K})\simeq9.25$ to $9.27$, corresponding to approximately $1.8\times10^9$~K, as shown in the right panel of Figure~\ref{fig:global_central_evolution}.

At these temperatures, Ne burning is initiated by the photodisintegration reaction $^{20}{\rm Ne}(\gamma,\alpha)^{16}{\rm O}$ rather than by the direct fusion of two heavy nuclei. The released alpha particles are subsequently captured by the remaining $^{20}{\rm Ne}$ through $^{20}{\rm Ne}(\alpha,\gamma)^{24}{\rm Mg}$, while further alpha capture can continue the processing toward $^{28}{\rm Si}$ \citep{Heger2000,Griffiths2025}. The strong temperature sensitivity of this reaction sequence helps explain why substantial central processing occurs over only days despite the narrow range of temperatures reached during this phase.

The composition evolution of the $1000\,M_\odot$ model shows this sequence directly. At central Ne depletion the centre remains strongly O rich with $X_{\rm c}(^{16}{\rm O})=0.842$, while $X_{\rm c}(^{24}{\rm Mg})=0.102$ and $X_{\rm c}(^{28}{\rm Si})=0.049$. Neon burning therefore redistributes the products inherited from C burning toward O, Mg, and the first significant Si without yet producing the much stronger compositional transformation associated with core O burning.

No model in the sequence develops a formal central convective core during Ne burning, although this does not exclude convection in detached burning shells. Radiative Ne burning has also been found in VMS calculations with large CO cores \citep{Yusof2013}, whereas lower-mass massive star models can develop convective Ne burning cores when nuclear heating exceeds the local energy losses. In the present models the absence of central convection is consistent with the rapidly increasing importance of neutrino cooling during the advanced evolution, although establishing the detailed balance would require a direct comparison of the local nuclear and neutrino energy generation rates.

Overall, core Ne burning marks the transition to an extremely rapid photodisintegration driven phase in which the O rich centre is processed toward Mg and Si while the formal central region remains radiative.

\paragraph{Core oxygen burning}

The interval from Ne to O exhaustion contracts further to between $0.593$ and $6.28$~d. The models reach $\log(T_{\rm c}/{\rm K})\simeq9.44$ - $9.47$, while the central density spans approximately $\log(\rho_{\rm c}/{\rm g\,cm^{-3}})=5.87$ - $6.60$. Oxygen burning is extremely temperature sensitive and proceeds mainly through fusion of $^{16}{\rm O}$, with the subsequent reaction flow building nuclei around the Si group. At these temperatures the rapid rise of the nuclear reaction rate competes with strong thermal neutrino losses, which contributes to the very short duration of this phase.

The chemical transformation is particularly clear in the $1000\,M_\odot$ model. At Ne exhaustion the centre is still strongly O rich with $X_{\rm c}(^{16}{\rm O})=0.842$, while $X_{\rm c}(^{24}{\rm Mg})=0.102$ and $X_{\rm c}(^{28}{\rm Si})=0.049$. By O exhaustion the central O abundance has fallen to $3.5\times10^{-7}$ and the centre contains $^{28}{\rm Si}$, $^{32}{\rm S}$, $^{36}{\rm Ar}$, and $^{40}{\rm Ca}$ with mass fractions of $0.540$, $0.203$, $0.028$, and $0.228$, respectively. Oxygen burning therefore produces a broad Si group composition rather than a centre dominated by Si alone.

The central convective region remains remarkably small during this phase. Its maximum extent is only $1.1$ to $2.2\,M_\odot$ from the $500$ to $900\,M_\odot$ models, while the $1000\,M_\odot$ model has no formal central convective O burning core. This does not exclude convection in surrounding burning shells, but it shows that the final central processing occupies only a very small fraction of stars whose earlier H and He burning cores contained several hundred solar masses. By O burning the evolutionary timescale has contracted to days or less, so local convective mixing can remain efficient within individual burning regions while large scale chemical redistribution can no longer keep pace with the evolution. The mixing timescale analysis in Section~\ref{sec:mixing_freezeout} shows that the surviving coupling becomes increasingly localized and is maintained primarily by convection, while rotational transport becomes too slow to reconnect the different layers during the final burning stages. Overall, O burning marks the rapid conversion of the O rich centre into Si group material while the active central region becomes extremely compact compared with the extended stellar envelope.

\subsubsection{Core Si burning}

The final measured nuclear interval is only $1.21$~hr in the $1000\,M_\odot$ model and $4.19$~hr in the $900\,M_\odot$ model. The longest value is $32.2$~hr for the $600\,M_\odot$ calculation. The $800\,M_\odot$ sequence does not reach central Si exhaustion. Across the late Si burning trajectories in Figure~\ref{fig:global_central_evolution}, the central conditions span approximately $\log(T_{\rm c}/{\rm K})\simeq9.6$ to $9.7$ and $\log(\rho_{\rm c}/{\rm g\,cm^{-3}})\simeq6.6$ to $8.1$.

The central convective region remains below $0.9\,M_\odot$ in every model that reaches Si exhaustion. At these temperatures Si burning proceeds through photodisintegration and light particle captures rather than through direct fusion of two Si nuclei. The reaction flow approaches quasi statistical equilibrium, and GeValNet25 represents this evolution using groups centred around $^{28}{\rm Si}$ and $^{56}{\rm Ni}$ with $^{44}{\rm Ti}(\alpha,\gamma)^{48}{\rm Cr}$ providing the connection between them. In the $1000\,M_\odot$ model, the adopted Si exhaustion point is reached when the normalized central composition is already about $99$\% iron group material. Electron captures within this material also begin to reduce $Y_e$ and remove energy through neutrinos, promoting continued contraction during the final hydrostatic evolution.

This contraction is accompanied by rapid central spin up. In the $1000\,M_\odot$ model $\Omega_{\rm c}$ reaches $3.99\times10^{-3}\ {\rm s^{-1}}$ at central Si exhaustion while the surface value remains only $5.25\times10^{-8}\ {\rm s^{-1}}$. Si burning therefore occurs within an extremely compact and rapidly evolving central region that is strongly separated both structurally and rotationally from the extended envelope.

The stippled area in the right panel of Figure~\ref{fig:global_central_evolution} shows the reference equation of state domain with $\Gamma_1<4/3$. This threshold is physically important because $\Gamma_1$ measures the pressure response to adiabatic compression, and in a radiation pressure dominated self gravitating region a value below $4/3$ indicates that this restoring response has become dynamically weak. Electron positron pair creation can produce such softening by converting thermal energy into particle rest mass \citep{Heger2002}. The central trajectories from $500$ through $900\,M_\odot$ remain outside this region, while the later $1000\,M_\odot$ trajectory reaches its boundary. Direct evaluation of the thermodynamic derivatives throughout the $1000\,M_\odot$ model nevertheless shows that $\Gamma_1$ never falls below $4/3$ in any mass shell. The smallest value over the complete evolution is $\Gamma_{1,\min}=1.33838$ during core He burning and occurs in the outer envelope at $M_r=998.76\,M_\odot$ and $r/R_\star=0.853$.

At central Si exhaustion the minimum $\Gamma_{1,\min}=1.34082$ and occurs near the processed core boundary at $M_r=193.24\,M_\odot$ and $r/R_\star=1.23\times10^{-3}$. This layer has $\log(T/{\rm K})=8.241$, $\log(\rho/{\rm g\,cm^{-3}})=0.924$, and a gas pressure fraction of only $\beta=0.045$. The centre instead has $\Gamma_1=1.35452$, while the pressure volume weighted value is $\left\langle\Gamma_1\right\rangle_{P{\rm d}V}=1.34765$. The late minimum therefore traces a radiation pressure dominated structural interface rather than a pair unstable central core. The $1000\,M_\odot$ model thus approaches the thermodynamic conditions associated with pair softening but does not develop a locally pair unstable core during the hydrostatic evolution followed here. The apparent contact of the central trajectory with the reference domain should therefore not be interpreted as the onset of pair instability.

\paragraph{Final hydrostatic structures}

The final points in the central density and temperature plane do not correspond to an identical nuclear stage because several calculations were continued farther than others. The $1000\,M_\odot$ model reaches $\log(T_{\rm c}/{\rm K})=10.027$ and $\log(\rho_{\rm c}/{\rm g\,cm^{-3}})=8.691$ because it was evolved substantially beyond the common Si exhaustion criterion. Stage matched structural comparisons are therefore considered separately below.

Table~\ref{tab:final_inventory} gives the integrated elemental masses in the final supplied hydrostatic structures. The models remain dominated by He, while the retained H inventory spans approximately $52$ - $130\,M_\odot$. The O inventory varies from $23.6\,M_\odot$ in the $600\,M_\odot$ model to $125.0\,M_\odot$ at $1000\,M_\odot$.

The advanced element inventory changes even more strongly across the sequence. The adopted Si group ranges from $2.39\,M_\odot$ in the $600\,M_\odot$ model to $52.49\,M_\odot$ at $1000\,M_\odot$, while the Fe group ranges from $0.54$ - $17.00\,M_\odot$. The $1000\,M_\odot$ structure alone contains $25.58\,M_\odot$ of Si and $16.21\,M_\odot$ of Fe. These quantities describe the material present in each final hydrostatic model rather than net nucleosynthetic yields, and the detailed Fe group partition, i.e. the isotopic distribution remains approximate within the advanced network.

\begin{table}[t]
\centering
\caption{
Evolutionary properties of the $v/v_{\rm crit}=0.1$ models.
The first block lists ages at central H exhaustion and the elapsed times between successive central depletion stages.
The second block lists the maximum central convective core mass within each interval.
The third block lists the surface equatorial velocity at the end of each interval.
The final block gives the central H and He mass fractions at the first crossing of $\log(T_{\rm eff}/{\rm K})=4.6$.
H, He, C, Ne, and O exhaustion use a central depletion threshold of $10^{-4}$ after the relevant species becomes active.
Si exhaustion uses the advanced central $^{28}$Si abundance with a post peak threshold of $10^{-3}$.
The $800\,M_\odot$ model does not reach this Si criterion.
}
\label{tab:evolution_stages}

\scriptsize
\setlength{\tabcolsep}{2.2pt}
\renewcommand{\arraystretch}{1.08}

\begin{tabular}{@{}lrrrrrr@{}}
\hline
\hline
 & 500 & 600 & 700 & 800 & 900 & 1000 \\
\multicolumn{7}{c}{Evolutionary time intervals} \\
\hline
H age [Myr] & 2.1225 & 1.9782 & 1.8809 & 1.8065 & 1.7506 & 1.7010 \\
He [kyr]    & 268.70 & 307.42 & 283.30 & 280.67 & 268.60 & 241.45 \\
C [kyr]     & 7.687  & 3.620  & 8.442  & 3.360  & 4.578  & 3.156 \\
Ne [d]      & 3.972  & 10.563 & 5.140  & 3.597  & 1.886  & 0.282 \\
O [d]       & 2.673  & 6.279  & 2.949  & 2.541  & 1.656  & 0.593 \\
Si [hr]     & 5.961  & 32.22  & 8.560  & \nodata & 4.193  & 1.211 \\
\hline
\multicolumn{7}{c}{Maximum central convective core mass [$M_\odot$]} \\
\hline
H  & 466.4 & 562.9 & 659.4 & 756.0 & 852.4 & 950.2 \\
He & 250.2 & 298.5 & 349.6 & 396.3 & 466.5 & 477.4 \\
C  & 50.88 & 24.26 & 43.86 & 47.04 & 68.69 & 143.45 \\
Ne & 0.00  & 0.00  & 0.00  & 0.00  & 0.00  & 0.00 \\
O  & 2.20  & 1.13  & 1.68  & 1.12  & 1.38  & 0.00 \\
Si & 0.65  & 0.34  & 0.68  & \nodata & 0.65 & 0.88 \\
\hline
\multicolumn{7}{c}{Surface equatorial velocity [${\rm km\,s^{-1}}$]} \\
\hline
H  & 170  & 194  & 222  & 239  & 273  & 289 \\
He & 25.5 & 3.88 & 2.10 & 21.3 & 45.0 & 13.7 \\
C  & 54.0 & 6.62 & 1.44 & 24.4 & 107  & 29.2 \\
Ne & 54.3 & 6.63 & 1.45 & 24.5 & 105  & 29.3 \\
O  & 52.8 & 6.63 & 1.48 & 27.4 & 100  & 76.9 \\
Si & 57.8 & 6.64 & 2.28 & \nodata & 106 & 130 \\
\hline
\multicolumn{7}{c}{First crossing of $\log(T_{\rm eff}/{\rm K})=4.6$} \\
\hline
$X_{\rm c}({\rm H})$  & 0.000 & 0.000 & 0.000 & 0.109 & 0.134 & 0.146 \\
$X_{\rm c}({\rm He})$ & 0.920 & 0.987 & 0.984 & 0.891 & 0.866 & 0.854 \\
\hline
\hline
\end{tabular}

\vspace{2pt}

\footnotesize
For the $1000\,M_\odot$ model with $v/v_{\rm crit}=0.4$, H exhaustion occurs at $1.7505$~Myr and the following He interval lasts $217.95$~kyr. Its maximum H and He burning convective core masses are $924.8$ and $779.1\,M_\odot$, while the corresponding endpoint surface velocities are $117$ and $4.36~{\rm km\,s^{-1}}$. It first crosses $\log T_{\rm eff}=4.6$ at $X_{\rm c}({\rm H})=0.043$. At H exhaustion, the surface H mass fractions are $0.751$ and $0.273$ for the $v/v_{\rm crit}=0.1$ and $0.4$ models, respectively.
\end{table}

\begin{table}[t]
\centering
\caption{
Elemental mass inventory of the final supplied hydrostatic structures for the six models with $v/v_{\rm crit}=0.1$.
All entries are in $M_\odot$.
$M_{\rm final}$ gives the stellar mass from the structure header.
Isotopes of each element are summed before integration over enclosed mass.
The Si group contains Si, S, Ar, Ca, and Ti.
The Fe group contains Cr, Fe, Co, and Ni.
The physical advanced $^{28}$Si channel is used and the legacy Si channel is excluded.
The $800\,M_\odot$ entry corresponds to its final available structure.
}
\label{tab:final_inventory}

\scriptsize
\setlength{\tabcolsep}{2.4pt}
\renewcommand{\arraystretch}{1.08}

\begin{tabular}{@{}lrrrrrr@{}}
\hline
\hline
$M_{\rm ini}$ & $M_{\rm final}$ & H & He & C & N & O \\
\hline
500  & 499.814 & 66.198  & 323.835 & 4.480 & 40.050 & 46.527 \\
600  & 594.144 & 52.422  & 457.848 & 4.949 & 45.948 & 23.576 \\
700  & 691.184 & 54.711  & 499.519 & 5.018 & 62.789 & 48.574 \\
800  & 789.434 & 85.965  & 566.225 & 4.557 & 67.642 & 47.230 \\
900  & 885.093 & 65.922  & 621.304 & 5.820 & 77.812 & 85.781 \\
1000 & 979.832 & 129.726 & 577.868 & 3.628 & 62.207 & 124.956 \\
\hline

\multicolumn{7}{c}{Intermediate mass elements} \\
\hline
$M_{\rm ini}$ & F & Ne & Na & Mg & Al & Si \\
\hline
500  & 0.000 & 7.111 & 0.000 & 3.241 & 0.000 & 2.488 \\
600  & 0.000 & 5.292 & 0.001 & 0.818 & 0.000 & 1.321 \\
700  & 0.000 & 8.463 & 0.307 & 4.005 & 0.008 & 2.311 \\
800  & 0.000 & 7.966 & 0.001 & 2.482 & 0.000 & 4.161 \\
900  & 0.000 & 7.916 & 0.002 & 6.980 & 0.000 & 6.220 \\
1000 & 0.000 & 4.288 & 0.000 & 7.669 & 0.000 & 25.581 \\
\hline

\multicolumn{7}{c}{Advanced element inventory} \\
\hline
$M_{\rm ini}$ & S & Ar & Ca & Ti & Cr & Fe \\
\hline
500  & 0.762  & 0.087 & 0.621  & 0.000 & 0.002 & 4.042 \\
600  & 0.483  & 0.077 & 0.507  & 0.000 & 0.001 & 0.768 \\
700  & 0.712  & 0.088 & 0.493  & 0.000 & 0.002 & 3.832 \\
800  & 1.457  & 0.197 & 1.006  & 0.000 & 0.001 & 0.522 \\
900  & 1.890  & 0.265 & 1.500  & 0.000 & 0.004 & 3.389 \\
1000 & 12.611 & 2.047 & 12.247 & 0.004 & 0.065 & 16.209 \\
\hline

\multicolumn{5}{c}{Grouped advanced inventory} \\
\hline
$M_{\rm ini}$ & Co & Ni & Si group & Fe group \\
\hline
500  & 0.154 & 0.217 & 3.958  & 4.414 \\
600  & 0.048 & 0.083 & 2.389  & 0.900 \\
700  & 0.169 & 0.184 & 3.605  & 4.187 \\
800  & 0.015 & 0.006 & 6.821  & 0.544 \\
900  & 0.178 & 0.110 & 9.873  & 3.680 \\
1000 & 0.656 & 0.070 & 52.490 & 17.000 \\
\hline
\hline
\end{tabular}
\end{table}

\subsection{Chemical and structural assembly of the advanced interior}
\label{sec:advanced_interior}

Figure~\ref{fig:global_central_evolution} shows how the full model sequence evolves at the surface and in the central $\rho_{\rm c}$ and $T_{\rm c}$ plane. We now focus on the $1000\,M_\odot$ model with $v/v_{\rm crit}=0.1$ to examine how this evolution builds the advanced stellar interior. The left panel of Figure~\ref{fig:advanced_interior} follows the central composition as a function of the time remaining until the final hydrostatic model, while the right panel shows where the corresponding burning and transport regions develop in radius.

The left panel first illustrates the enormous contraction of the nuclear evolutionary clock. Central H exhaustion occurs about $2.45\times10^{5}$~yr before the final model, whereas central He exhaustion occurs only about $3.16\times10^{3}$~yr before it. The long preceding He burning phase also establishes the composition inherited by the advanced evolution. By He exhaustion the centre is already O rich, with $X_{\rm c}(^{16}{\rm O})=0.780$, while the remaining $^{20}{\rm Ne}$, $^{12}{\rm C}$, and $^{24}{\rm Mg}$ mass fractions are $0.114$, $0.063$, and $0.043$. The advanced stages therefore begin from a centre whose composition has already been largely set by He burning.

Moving from left to right through the late part of the left panel shows the successive depletion of C, Ne, O, and Si over increasingly compressed intervals in time. The reaction physics responsible for these changes was discussed in Section~\ref{sec:global_central_evolution}. Here, the important point is how rapidly the composition evolves from an O rich state into the Si and Fe groups once the advanced sequence begins. The $^{28}{\rm Si}$ abundance reaches a maximum of $0.869$ only about $16$~min before the final model and is then rapidly consumed. Central Si exhaustion is reached about $3.2$~min before the endpoint, when approximately $99$\% of the normalized central composition belongs to the Fe group.

The individual Fe group curves in the same panel show rapid redistribution among the tracked nuclei over these final minutes. In GeValNet25 this evolution includes the effective weak interaction sequence through Ni, Co, Fe, and Cr that approximates the decrease in $Y_e$. The detailed partition among individual Fe group isotopes should therefore not be interpreted as precision nucleosynthesis. Their summed abundance instead provides the more useful measure of how completely the central material has been processed by the end of Si burning.

The right panel of Figure~\ref{fig:advanced_interior} places this rapidly changing central composition into its radial context. The horizontal axis again gives $t_{\rm end}-t$, where $t_{\rm end}$ is the age of the final hydrostatic GENEC model rather than the time of hydrodynamic collapse.  Reading the right panel from early to late times shows that each new burning stage appears progressively closer to the endpoint and within a progressively more compact part of the star. The first C burning region appears about $2.90\times10^{3}$~yr before the final model. Ne and O burning first appear only $3.36$ and $2.67$~d before the endpoint, followed by Si burning about $4.75$~hr before it. An Fe group core becomes identifiable only $5.92$~min before the final model. These values mark the first appearance anywhere in the star of a region undergoing the corresponding burning stage and are therefore distinct from the central fuel exhaustion intervals discussed above and listed in Table~\ref{tab:evolution_stages}.

The dashed isomass curves in the right panel show how strongly the interior contracts while the already expanded envelope remains at radii of several thousand solar radii through the advanced stages. In the final structure the inner $100\,M_\odot$ is contained within only $0.088\,R_\odot$, while the $200\,M_\odot$ coordinate lies at $2.12\,R_\odot$. Even the $900\,M_\odot$ coordinate remains within about $126\,R_\odot$, whereas the photospheric radius reaches several thousand solar radii. The dotted isotherms reveal a similar thermal concentration. At the final stage the $\log(T/{\rm K})=9$ surface lies near $0.16\,R_\odot$ and encloses about $163\,M_\odot$, while the $\log(T/{\rm K})=8$ surface lies near $0.96\,R_\odot$ and encloses about $195\,M_\odot$.

\begin{figure}[t]
    \centering
    \includegraphics[width=1.0\columnwidth]{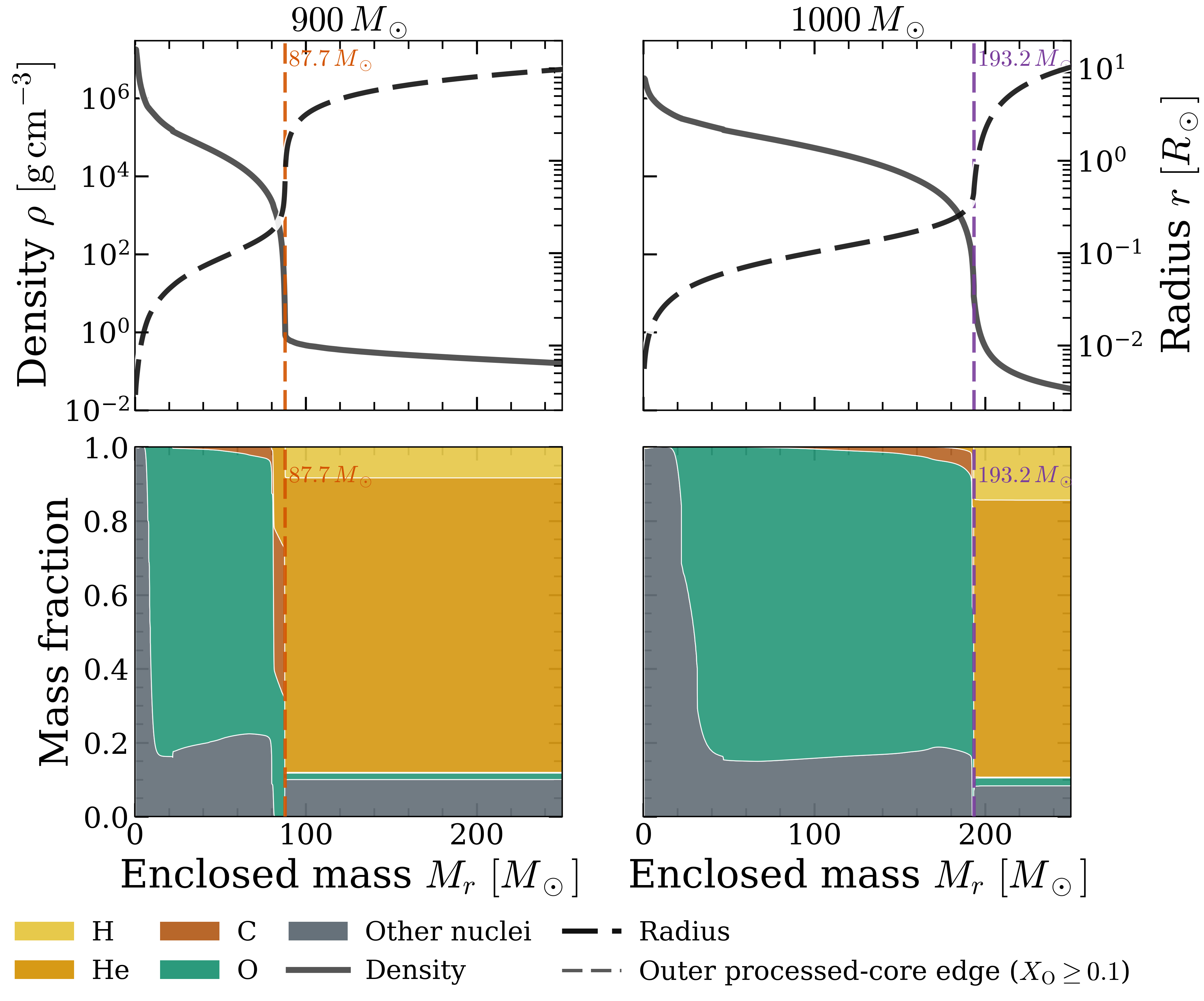}
    
    \caption{
    Internal structure of the $900\,M_\odot$ and $1000\,M_\odot$ models at central Si exhaustion.
    The left and right columns show the $900\,M_\odot$ and $1000\,M_\odot$ models over the inner $250\,M_\odot$, respectively.
    The upper panels show density as the solid grey curve using the left vertical axis and radius as the black dashed curve using the right vertical axis.
    The lower panels show the stacked composition fractions of H, He, C, O, and the grouped remaining nuclei as functions of enclosed mass.
    The vertical dashed lines mark the outer boundary where $X_{\rm O}\geq0.1$ and identify enclosed masses of $87.7\,M_\odot$ and $193.2\,M_\odot$ for the two models.
    }
    \label{fig:terminal_structure}
\end{figure}

The hatching and stippling in the right panel also show that the transport structure becomes increasingly layered as the nuclear timescale shortens. Broad convective regions associated with the earlier burning phases give way to alternating convective and radiative layers during the advanced evolution, while the Fe group material remains confined to the innermost region. The two panels of Figure~\ref{fig:advanced_interior} therefore show the same physical transition from complementary perspectives. Whether or not these layers can exchange material depends on the mixing timescale rather than on the convective or radiative structure alone. Section~\ref{sec:mixing_freezeout} quantifies this directly by comparing the local mixing time with the remaining evolutionary time and by separating the contributions from convection and rotational transport. The left panel demonstrates the rapid chemical processing of the centre, while the right panel shows how this processing becomes confined to an increasingly compact and layered interior.

\subsection{Terminal core structure at central Si exhaustion}
\label{sec_terminal_structure}

Figure~\ref{fig:terminal_structure} shows the inner structures of the $900$ and $1000\,M_\odot$ models at the same central Si exhaustion criterion. In the bottom panels of Figure~\ref{fig:terminal_structure}, the outer boundary with $X_{\rm O}\geq0.1$ is at $87.7\,M_\odot$ in the $900\,M_\odot$ model and at $193.2\,M_\odot$ in the $1000\,M_\odot$ model. The mass extent of the O rich processed interior therefore increases by a factor of about $2.2$ even though the initial stellar mass changes by only about $11$\%. In both models this interface separates the strongly processed O rich interior from layers that rapidly become dominated by He.

The same boundary is independently visible in the structural quantities shown in the top panels of Figure~\ref{fig:terminal_structure}. Across the $87.7\,M_\odot$ interface of the $900\,M_\odot$ model, the density decreases by about $3.27$ dex while the radius increases by about $1.05$ dex. The $1000\,M_\odot$ model shows corresponding changes of about $2.89$ dex in density and $0.79$ dex in radius near $193.2\,M_\odot$. The abundance transition therefore coincides with the boundary between the compact processed core and the much more extended He rich layers. This behavior follows directly from the mass continuity equation,
\begin{equation}
\frac{{\rm d}r}{{\rm d}m}
=
\frac{1}{4\pi r^2\rho},
\end{equation}
where $r$ is the radial coordinate, $m$ is the enclosed mass, and $\rho$ is the local density. A sharp decrease in $\rho$ therefore increases ${\rm d}r/{\rm d}m$, so that a relatively small increase in enclosed mass spans a much larger radial interval.

The physical origin of the different interface locations can be traced to the conditions established before the final burning stages. As seen in the right panel of Figure~\ref{fig:global_central_evolution}, the $900$ and $1000\,M_\odot$ central trajectories begin to separate during the advanced evolution following He exhaustion. At this stage the $900\,M_\odot$ model retains $X_{\rm c}(^{12}{\rm C})=0.100$, while the $1000\,M_\odot$ model retains only $0.063$. Their central states also differ, with $\log(T_{\rm c}/{\rm K})=8.573$ and $\log(\rho_{\rm c}/{\rm g\,cm^{-3}})=2.933$ for the $900\,M_\odot$ model compared with $8.585$ and $2.768$ for the $1000\,M_\odot$ model.

These differences are important because the carbon burning rate depends approximately on
\begin{equation}
\epsilon_{\rm C}
\propto
\rho X_{\rm C}^{2}
\langle\sigma v\rangle .
\end{equation}
Here $\epsilon_{\rm C}$ is the specific energy generation rate from carbon burning, $\rho$ is the local density, $X_{\rm C}$ is the carbon mass fraction, and $\langle\sigma v\rangle$ is the thermally averaged reaction rate for carbon fusion. The quadratic dependence on $X_{\rm C}$ makes the residual carbon abundance left by core He burning particularly important for the subsequent evolution. The $1000\,M_\odot$ model therefore enters C burning with less carbon fuel and a lower central density, while its slightly higher temperature partly compensates through the strong temperature sensitivity of carbon fusion. The resulting contraction and C burning history changes the conditions under which the subsequent C and O burning regions develop. Similar sensitivity to the carbon abundance left by He burning and to the later C and O shell evolution is well established in classical massive star calculations \citep{Sukhbold2014,Limongi2018,Patton2020,Griffiths2025}. In the present models this sensitivity is amplified to much larger mass scales and produces a shift of more than $100\,M_\odot$ in the boundary of the processed interior by central Si exhaustion.

\subsection{Rotation and angular momentum evolution}
\label{sec_rotation}

\begin{figure*}[t]
    \centering
    \includegraphics[width=0.98\textwidth]{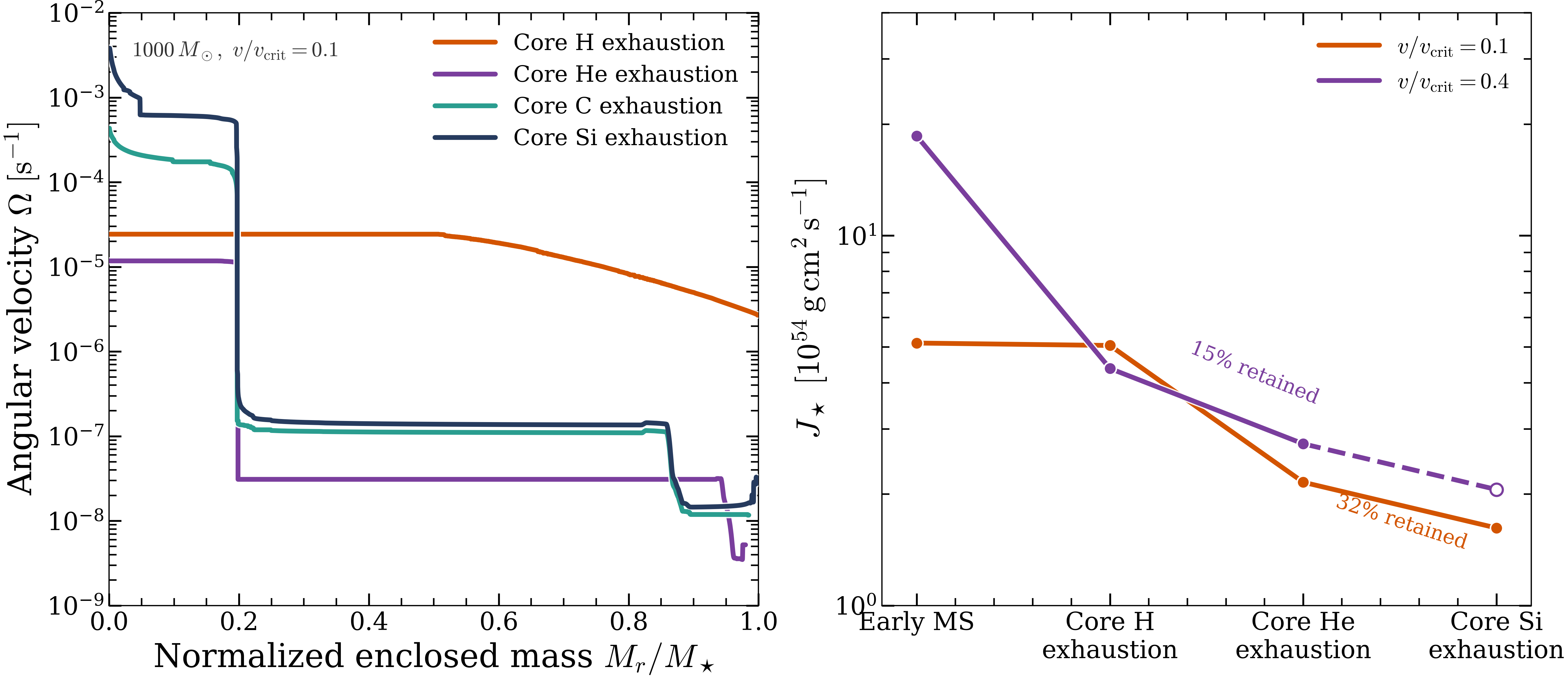}

    \caption{
    Rotational evolution of the $1000\,M_\odot$ models.
    The left panel shows the angular velocity $\Omega$ as a function of normalized enclosed mass for the $v/v_{\rm crit}=0.1$ model at core H, He, C, and Si exhaustion.
    The right panel shows the total stellar angular momentum $J_\star$ at the early main sequence and at subsequent core exhaustion stages for models with $v/v_{\rm crit}=0.1$ and $0.4$.
    The labelled percentages give the fraction of the early main sequence angular momentum retained at the indicated stages.
    The $v/v_{\rm crit}=0.4$ calculation was evolved only to core He exhaustion.
    The dashed purple segment therefore shows a conditional projection to core Si exhaustion obtained by applying the He to Si angular momentum retention factor measured in the $v/v_{\rm crit}=0.1$ model.
    }
    \label{fig_rotation}
\end{figure*}

Figure~\ref{fig_rotation} shows the rotational evolution of the $1000\,M_\odot$ models. The left panel follows the angular velocity profile of the $v/v_{\rm crit}=0.1$ model from core H exhaustion to core Si exhaustion, while the right panel shows how the total stellar angular momentum evolves for the two initial rotation rates.

The left panel of Figure~\ref{fig_rotation} shows that the angular velocity profile changes substantially as the star progresses through successive burning stages. At core H exhaustion the inner $\sim 0.5 \, M_\star$ has a nearly flat profile with $\Omega\simeq2.4\times10^{-5}\ {\rm s^{-1}}$, followed by a gradual decline through the outer layers. By core He exhaustion the central angular velocity has decreased to $\simeq1.17\times10^{-5}\ {\rm s^{-1}}$ and the nearly uniform inner region extends only to about $0.2 \, M_\star$. During these comparatively long evolutionary phases, meridional circulation, shear transport, and efficient mixing within convective regions have sufficient time to redistribute angular momentum and smoothen large gradients in $\Omega$ \citep{Maeder2000,Meynet2000}.

The behavior changes after core He exhaustion. At core C exhaustion the central angular velocity has increased to $\simeq4.32\times10^{-4}\ {\rm s^{-1}}$, while a steep gradient develops across the inner $\sim 0.2 \, M_\star$. By core Si exhaustion the central value reaches $\simeq3.99\times10^{-3}\ {\rm s^{-1}}$, corresponding to an increase by a factor of about $340$ from He exhaustion. This rapid spin up follows the strong contraction of the advanced core discussed in the preceding sections. For shellular rotation the specific angular momentum can be written approximately as
\begin{equation}
\bar{j}=\frac{2}{3}r^2\Omega ,
\end{equation}
where $\bar{j}$ is the shell averaged specific angular momentum, $r$ is the radial coordinate, and $\Omega$ is the angular velocity. A contracting mass shell can therefore increase its angular velocity very strongly even when its specific angular momentum changes only modestly.

The shape of the late profiles is also closely connected to the internal structure. At Si exhaustion the sharp decrease in $\Omega$ near $M_r/M_\star\simeq0.2$ occurs at almost the same mass coordinate as the $193.2\,M_\odot$ processed core boundary identified in Figure~\ref{fig:terminal_structure}. The rotation profile therefore reflects the same separation between the compact advanced core and the extended envelope. The flatter portions farther out should not be interpreted as a single convective region. As shown by Figure~\ref{fig:advanced_interior}, the advanced star contains alternating convective and radiative layers associated with the different burning shells. Instead, much of the outer $\Omega$ profile retains the angular momentum distribution established during the earlier and much longer evolutionary phases. Once C, Ne, O, and Si burning begin, their rapidly decreasing nuclear timescales leave progressively less time for large scale angular momentum redistribution.

The right panel of Figure~\ref{fig_rotation} shows the complementary evolution of the total angular momentum. The $v/v_{\rm crit}=0.4$ model begins the main sequence with about $3.75$ times more angular momentum than the $v/v_{\rm crit}=0.1$ model, yet the two values have already become comparable by core H exhaustion. The more rapidly rotating model therefore loses a substantially larger fraction of its initial angular momentum during H burning. Its stronger initial rotation brings the surface closer to the $\Omega\Gamma$ limit, increasing the removal of mass and angular momentum as the star expands. By H exhaustion it has already lost about $34\,M_\odot$. Angular momentum transported toward the outer layers can be removed efficiently because the specific angular momentum of surface material scales approximately with $r^2\Omega$, so even modest surface mass loss can remove a disproportionately large amount of $J_\star$ \citep{Maeder2000,Yusof2013}. The $v/v_{\rm crit}=0.4$ model therefore does not preserve its much larger initial angular momentum reservoir.

By core He exhaustion the $v/v_{\rm crit}=0.4$ model retains only about $15$\% of its early main sequence angular momentum, meaning that approximately $85$\% has already been lost. The corresponding $v/v_{\rm crit}=0.1$ model retains about $42$\% at He exhaustion and about $32$\% by core Si exhaustion. Thus the slow model continues to lose global angular momentum during the advanced evolution even while its central angular velocity rises by more than two orders of magnitude. The dashed segment for the $v/v_{\rm crit}=0.4$ model is a conditional projection obtained by applying the measured He to Si retention factor of the $v/v_{\rm crit}=0.1$ model and does not represent an evolved advanced stage calculation. The central result of Figure~\ref{fig_rotation} is therefore that rapid contraction produces a strongly spinning inner core while the star as a whole continues to lose angular momentum, and that a higher initial rotation rate does not necessarily translate into a proportionally larger angular momentum reservoir at the end of the long H and He burning phases. The convergence of the two rotating models in total $J_\star$ should not be interpreted as convergence toward a nonrotating state, but instead reflects the preferential angular momentum loss of the initially more rapidly rotating model.

\subsection{Evolution of the global elemental inventory}
\label{sec:elemental_inventory}

\begin{figure}[t]
    \centering
    \includegraphics[width=0.98\columnwidth]{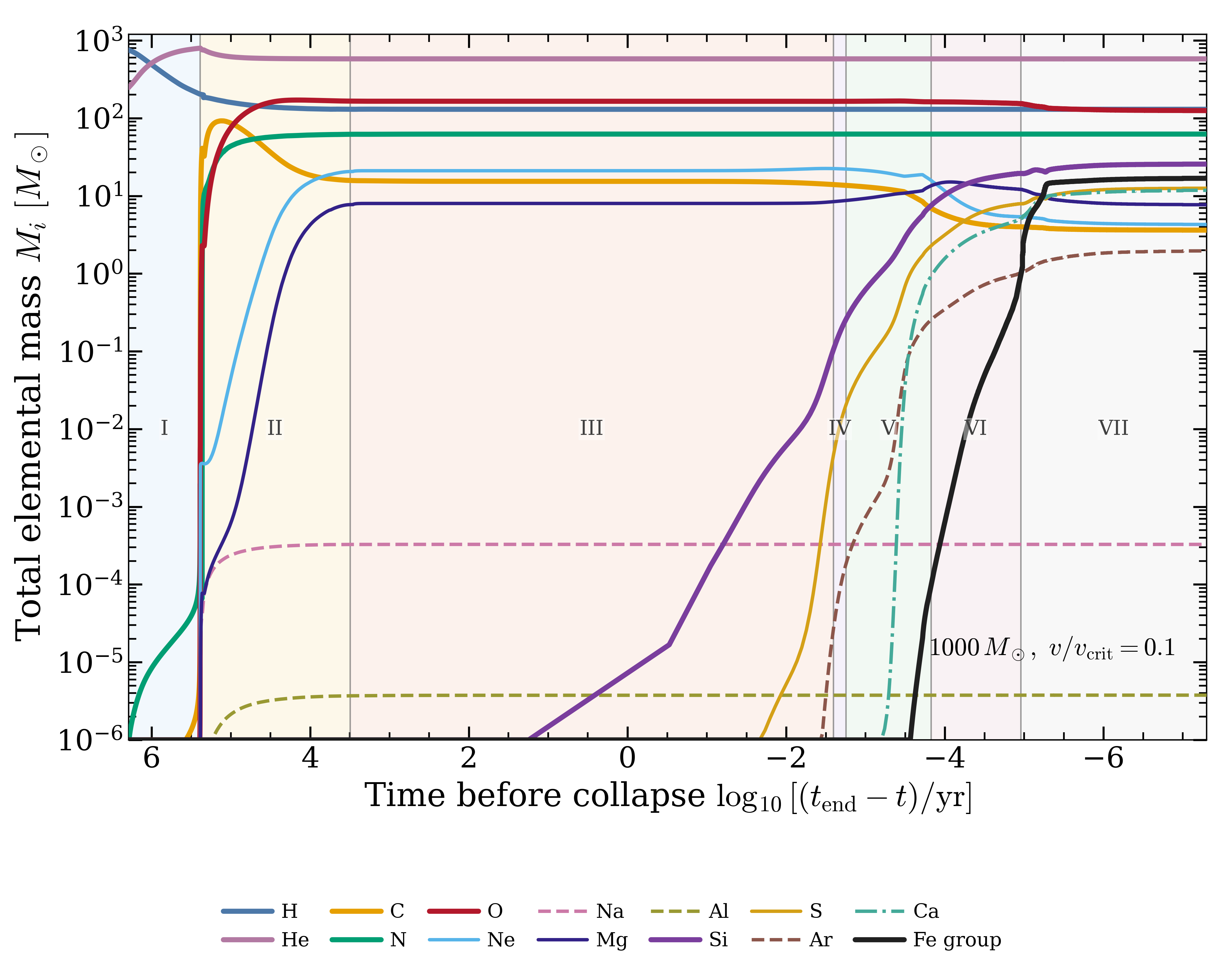}

    \includegraphics[width=0.98\columnwidth]{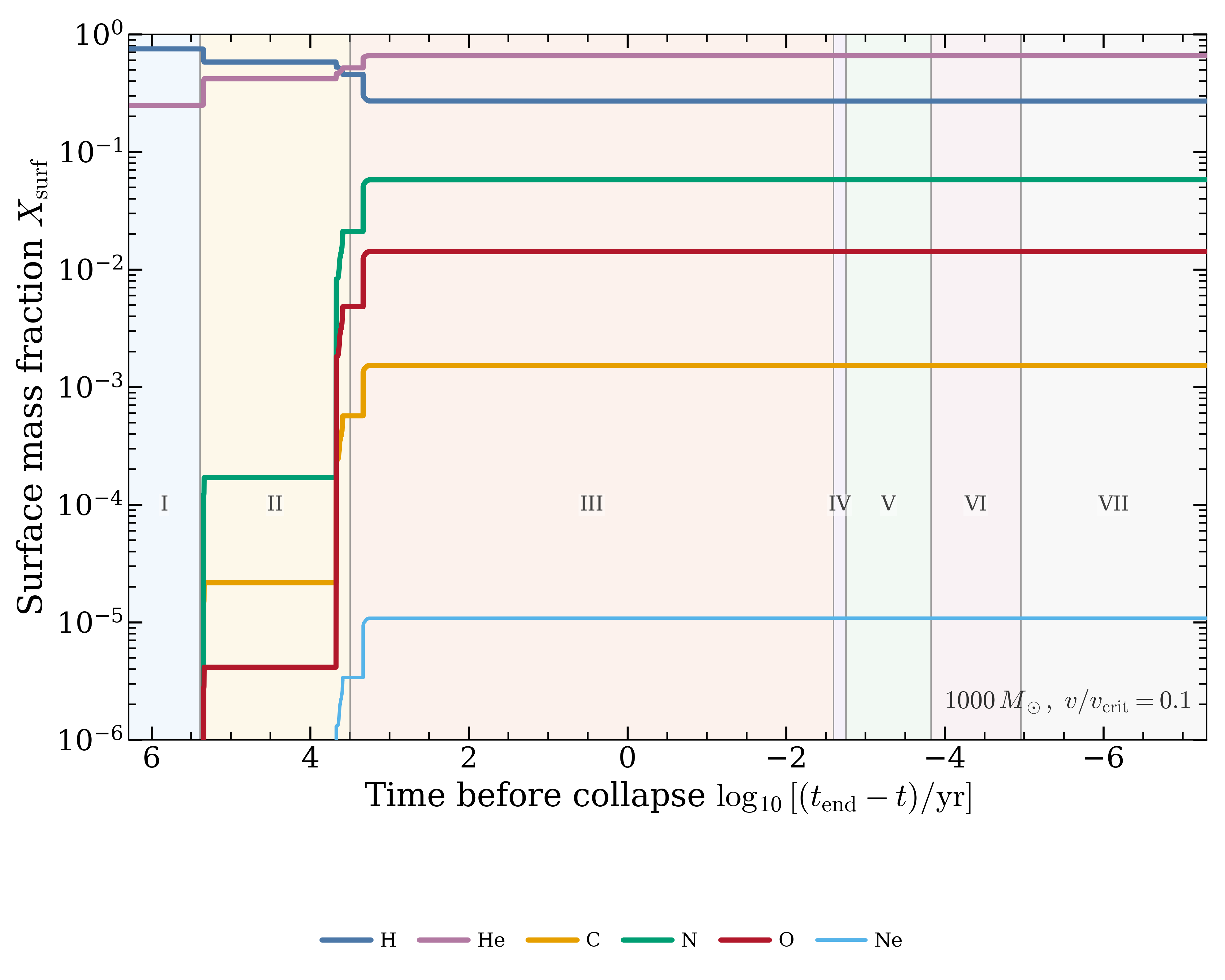}

    \caption{
    Chemical evolution of the $1000\,M_\odot$ model with $v/v_{\rm crit}=0.1$.
    The upper panel shows the total elemental masses contained within the star, obtained by integrating the corresponding isotope mass fractions over the stellar mass at each model snapshot.
    The Fe group contains the summed Cr, Fe, Co, and Ni inventories.
    The lower panel shows the corresponding surface elemental mass fractions.
    Only elements that reach $X_{\rm surf}\geq10^{-6}$ are shown in the lower panel.
    F, Na, Mg, Al, Si, S, Ar, Ca, Ti, and the Fe group remain below this threshold throughout the evolution.
    The shaded intervals denote I: core H burning, II: core He burning, III: core C burning, IV: core Ne burning, V: core O burning, VI: core Si burning, and VII: evolution after central Si exhaustion.
    Both panels use the same time coordinate measured relative to the final hydrostatic model.
    }
    \label{fig:elemental_inventory}
\end{figure}

Figure~\ref{fig:elemental_inventory} follows both the total elemental inventory and the surface composition of the $1000\,M_\odot$ model with $v/v_{\rm crit}=0.1$. The stellar mass decreases to $979.83\,M_\odot$ by the terminal model through the combined surface mass loss described in Section~\ref{sec:methods}, including the prescribed radiative winds and the rotationally regulated loss associated with the $\Omega\Gamma$ limit. We divide the evolution into stages according to the active central burning phase: I, H burning; II, He burning; III, C burning; IV, Ne burning; V, O burning; VI, Si burning; and VII, evolution after central Si exhaustion. During stage I, core H burning reduces the H inventory from $751.6$ to $203.3\,M_\odot$ while He increases from $248.4$ to $796.7\,M_\odot$. Heavy elements remain negligible at H exhaustion. This conversion occurs within the very large H-burning mixed region, whose convective core reaches about $950\,M_\odot$ in this model. The $v/v_{\rm crit}=0.4$ calculation terminates at core He exhaustion and is therefore not included in the advanced elemental inventory comparison.

Stage II produces the first major heavy-element reservoirs. Triple-$\alpha$ burning raises the total C mass to a maximum of $92.4\,M_\odot$, after which $^{12}{\rm C}(\alpha,\gamma)^{16}{\rm O}$ shifts an increasing fraction of the processed material toward O. The O inventory reaches $171.1\,M_\odot$ late in He burning and remains at $165.8\,M_\odot$ at He exhaustion, whereas C has fallen to $15.8\,M_\odot$. These large integrated masses reflect the broad He-burning core, which reaches about $477\,M_\odot$, together with the continued chemical transport operating in the rotating model.

The strongest consequence of this coupling is the production of primary N during stage II. Its total mass rises from $6.6\times10^{-4}\,M_\odot$ at H exhaustion to $62.1\,M_\odot$ at He exhaustion and changes by less than 1\% thereafter. Since the star begins without CNO nuclei, this reservoir requires freshly synthesized C and O to be transported into H-burning material where they provide the catalysts for CNO cycling and primary N production \citep{Nandal2024b}. In the present model this transport is supplied by the combined action of convection and rotational mixing during the $\simeq2.4\times10^{5}$ yr He burning phase. The process therefore depends on the local transport between burning regions rather than on the total stellar angular momentum alone, while the subsequent advanced stages are too short to redistribute comparable amounts of material.

The lower panel of Figure~\ref{fig:elemental_inventory} shows when this processed material becomes visible at the stellar surface. The first CNO enrichment appears around the transition from core H to core He burning, with N reaching $X_{\rm N}\simeq1.6\times10^{-4}$ while C and O remain considerably lower. This early N dominance is expected for a metal free VMS because high core temperatures allow trace $^{12}{\rm C}$ to be produced by triple-$\alpha$ reactions already during H burning, after which the CNO cycle rapidly converts much of the available C and O into $^{14}{\rm N}$ \citep{Tsiatsiou2024, Nandal2025}. The surface CNO abundances then remain nearly constant through much of stage II even though the internal C and O reservoirs continue to grow. This separation reflects the finite transport time between the He burning region and the surface, so freshly synthesized C and O can accumulate in the intervening layers and feed the H burning shell without immediately altering the photospheric composition. Near He exhaustion a second and much larger enrichment occurs at $\log_{10}[(t_{\rm end}-t)/{\rm yr}]\simeq3.5$. Rotating low metallicity models show that the progressive injection of C and O into the H shell strengthens CNO burning and can promote deeper convective and rotational coupling, allowing the previously accumulated CNO rich material to reach the surface \citep{Ekstrom2008,Murphy2021a}. N rises most strongly because part of the transported C and O is processed through the proton rich shell before reaching the envelope, while the simultaneous increases in C and O show that an increasing fraction of He burning material also reaches the surface without complete CNO processing. Ne appears only during this late enrichment and remains near $10^{-5}$, consistent with its secondary production from He burning material, including the conversion of primary $^{14}{\rm N}$ into $^{22}{\rm Ne}$. F, Na, Mg, Al, Si, S, Ar, Ca, Ti, and the Fe group remain below $X_{\rm surf}=10^{-6}$ and are therefore not shown.

Stage III marks the transition to advanced burning, but the whole-star composition initially changes only modestly. The broad mixed cores of H and He burning are replaced by the layered central and shell structure shown in Figure~\ref{fig:advanced_interior}, while the remaining evolutionary time contracts rapidly. The C inventory decreases from $15.8$ to $14.0\,M_\odot$ by C exhaustion, whereas most of the previously established O and N reservoirs remain nearly unchanged. Stage IV, corresponding to core Ne burning, is shorter still and produces little change in the global elemental inventory despite substantial central and shell burning. This reflects the small mass involved and the short remaining timescale rather than an absence of active shell burning.

The first pronounced change among the heavier species appears during stage V, when core O burning drives a clear rise in Mg and the Si-group elements. By central O exhaustion, Ne has decreased to $15.5\,M_\odot$ while Mg has grown to $13.7\,M_\odot$, and the star already contains $7.73\,M_\odot$ of Si and $2.34\,M_\odot$ of S. The star then contains $7.73\,M_\odot$ of Si, $2.34\,M_\odot$ of S, $0.26\,M_\odot$ of Ar, and $0.97\,M_\odot$ of Ca. At the same time the total O inventory remains as large as $162.4\,M_\odot$, even though central O has been essentially exhausted. Central fuel depletion therefore affects only the deepest processed region, while a much larger O-rich reservoir survives in the surrounding structure.

Stages VI and VII complete the processing toward the Si and Fe groups. At central Si exhaustion the star contains $21.2\,M_\odot$ of Si, $9.45\,M_\odot$ of S, $1.39\,M_\odot$ of Ar, $9.07\,M_\odot$ of Ca, and $9.06\,M_\odot$ of Fe-group material. During Si burning, the composition shifts from nuclei grouped around Si toward the Fe-group species represented in the reduced network, including Cr, Fe, Co, and Ni. Continued processing after central Si depletion raises the Fe-group inventory to $17.1\,M_\odot$ in the terminal model, while Si, S, Ar, and Ca reach $25.5$, $12.6$, $2.04$, and $12.2\,M_\odot$. The star nevertheless retains $124.5\,M_\odot$ of O and $62.4\,M_\odot$ of N, showing that the final advanced stages transform only the compact inner structure while preserving much of the chemical inventory established during the longer H- and He-burning phases. Therefore, the global chemical inventory is established primarily during the long H and He burning phases, most notably through the production of a large primary N reservoir and an O rich interior. The subsequent advanced stages strongly process the compact centre toward the Si and Fe groups, but are too short to substantially erase the much larger O and N reservoirs already present in the star.

\subsection{Composition of the cumulative wind ejecta}
\label{sec:wind_composition}

\begin{figure}[t]
    \centering
    \includegraphics[width=0.98\columnwidth]{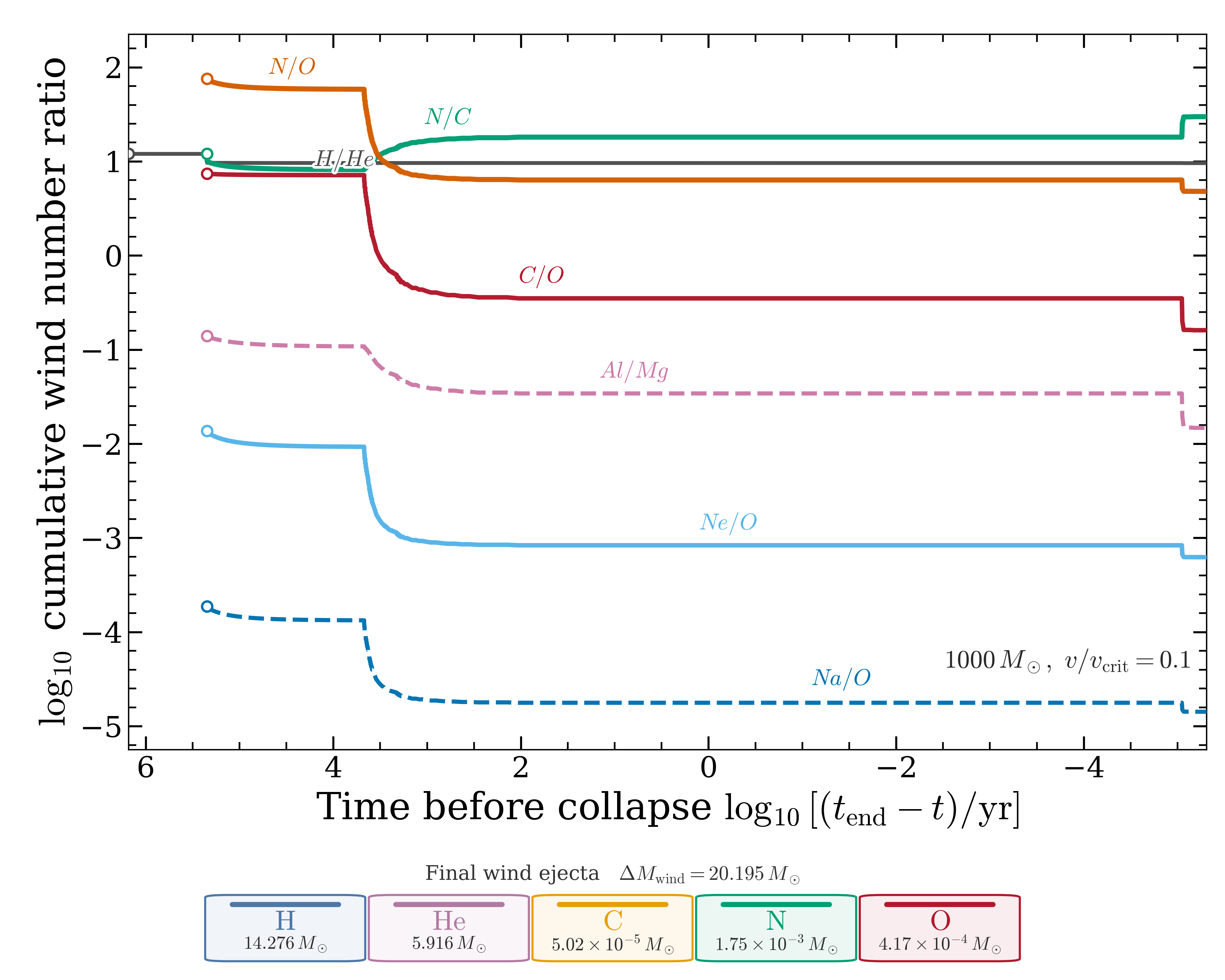}
    \caption{
    Evolution of the cumulative wind-ejecta composition for the
    $1000\,M_\odot$ model with $v/v_{\rm crit}=0.1$.
    For each decrease in the stellar mass, the lost material is assigned
    the mean surface composition across that evolutionary interval and
    added to the cumulative elemental ejecta.
    The curves show the resulting elemental
    number ratios, $\log_{10}[(M_i/A_i)/(M_j/A_j)]$, where $M_i$ is the
    cumulative wind-carried elemental mass and $A_i$ is its representative
    atomic weight.
    The ratios are raw model number ratios and are not normalized to a
    solar abundance pattern.
    Open circles mark the first plotted values for which both elements
    exceed the adopted cumulative-mass threshold.
    The time coordinate is measured relative to the final hydrostatic model. The boxes below the panel give the final cumulative masses of H, He, C, N, and O carried away by the wind over the computed evolution. Only these five elements are shown because the cumulative wind-ejected masses of the remaining tracked elements are below $10^{-6}\,M_\odot$ and are negligible on the scale of the total ejecta.
    }
    \label{fig:wind_ratios}
\end{figure}

The $1000\,M_\odot$ model loses $20.2\,M_\odot$ before reaching the terminal stage, and Figure~\ref{fig:wind_ratios} follows the cumulative composition of this material. The adopted mass loss combines the radiative prescriptions of \citet{Vink2001} and \citet{DeJager1988} with the rotational enhancement associated with the approach to the $\Omega\Gamma$ limit. At primordial metallicity the radiative contribution is expected to be weak, making rotation particularly important for removing mass from the surface \citep[for less massive Pop~III stars, also see][]{Liu_rot2021}.

Most of the mass is lost during H and He burning, which together account for almost the entire stellar lifetime. The cumulative loss already reaches about $20.0\,M_\odot$ during early He burning and $20.165\,M_\odot$ by He exhaustion, compared with $20.195\,M_\odot$ in the terminal model. The dependence on rotation is substantial: the $v/v_{\rm crit}=0.4$ model has already lost about $34\,M_\odot$ by H exhaustion, exceeding the total mass removed from the reference model over its entire computed evolution. This long H and He burning interval is also when most of the global angular momentum loss occurs, as shown in Section~\ref{sec_rotation}. The advanced burning phases therefore add almost no mass to the wind, and their nucleosynthetic products remain largely confined to the stellar interior.

The most prominent change in Figure~\ref{fig:wind_ratios} occurs during late core He burning near $\log_{10}[(t_{\rm end}-t)/{\rm yr}]\simeq3.5$. Over this interval the surface O mass fraction increases from $4.2\times10^{-6}$ to $4.8\times10^{-3}$, much more rapidly than C or N. Such surface CNO enrichment can in principle provide additional line opacity and strengthen radiative driving at very low initial metallicity \citep{Krt2009}, although the present calculation does not isolate this contribution from the adopted mass-loss prescription. The abundance change drives the simultaneous decline of $\mathrm{N}/\mathrm{O}$, $\mathrm{C}/\mathrm{O}$, $\mathrm{Ne}/\mathrm{O}$, and $\mathrm{Na}/\mathrm{O}$, all of which contain O in the denominator. The change is therefore not caused by core O burning, which occurs much later, but by the outward transport of increasingly O-rich material produced during He burning. Triple-$\alpha$ burning first supplies C, while continued $^{12}{\rm C}(\alpha,\gamma)^{16}{\rm O}$ progressively enriches the He-burning material in O. Rotational transport then carries part of this material toward the H-burning layers and surface, where C and O also provide the catalysts required for primary N production.

This processing leaves a clear CNO signature in the integrated wind. The final ratios reach $\log(\mathrm{N}/\mathrm{C})=1.48$ and $\log(\mathrm{N}/\mathrm{O})=0.68$, corresponding to $\mathrm{N}/\mathrm{C}$ and $\mathrm{N}/\mathrm{O}$ number ratios of about $30$ and $4.8$, respectively. In absolute terms, the cumulative wind ejecta contain $1.8\times10^{-3}\,M_\odot$ of N by the terminal model. At the same time $\log(\mathrm{C}/\mathrm{O})=-0.80$, so C remains substantially less abundant than O. The final ratios reach $\log(\mathrm{N}/\mathrm{C})=1.48$ and $\log(N/O)=0.68$, corresponding to N/C and N/O number ratios of about $30$ and $4.8$, respectively. At the same time $\log(C/O)=-0.80$, so C remains substantially less abundant than O. The wind is therefore strongly N enhanced rather than simply CNO enriched, directly linking the material removed from the star to the primary-N reservoir established during He burning in Figure~\ref{fig:elemental_inventory}. The nearly constant $\log(H/He)\simeq0.98$ further shows that the ejecta remain dominated by the H- and He-rich envelope rather than by material exposed from the advanced-burning core.

The heavier light-element ratios show a very different behavior. The final values of $\log(Ne/O)=-3.20$ and $\log(Na/O)=-4.85$ indicate that the material carried by the wind contains very little Ne or Na relative to O, while $\log(Al/Mg)=-1.83$ shows that Al remains strongly suppressed relative to Mg. These weak NeNa and MgAl signatures contrast with the substantial production of heavier elements inside the star shown in Figure~\ref{fig:elemental_inventory}. The modeled wind therefore primarily carries the products of H and He burning, especially primary N, while the products of the advanced burning stages remain buried in the compact interior. The very low $\mathrm{Ne}/\mathrm{O}$ ratio could therefore provide a chemical signature of enrichment dominated by such winds, although its appearance in a galaxy would depend on dilution and contributions from other enrichment sources. 

\subsection{Chemical accessibility under deeper stripping}
\label{sec:deeper_stripping}

\begin{figure}[t]
    \centering
    \includegraphics[width=0.98\columnwidth]{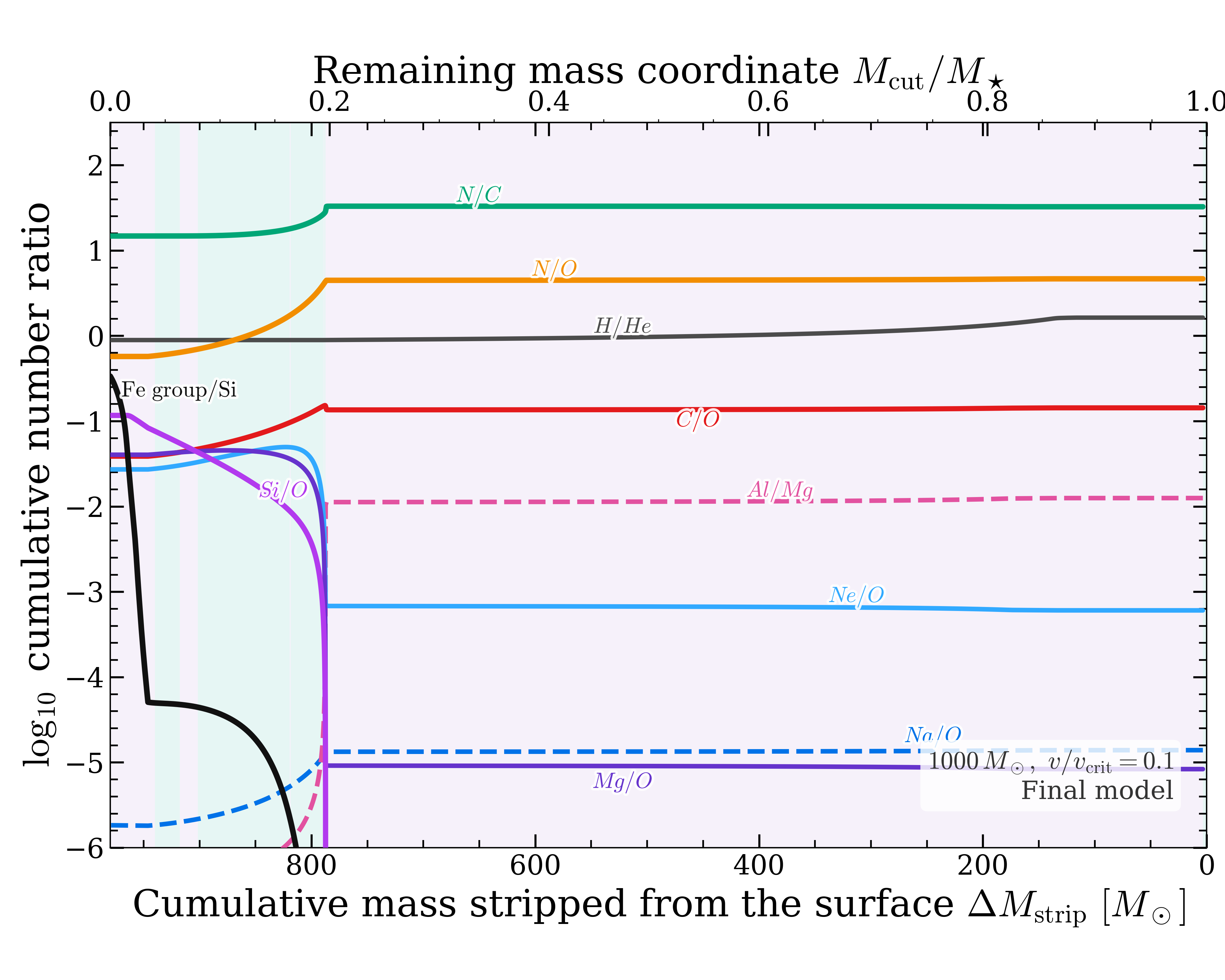}
    \caption{
    Conditional cumulative stripping experiment for the terminal
    $1000\,M_\odot$ model with $v/v_{\rm crit}=0.1$.
    The upper axis gives the remaining mass coordinate
    $M_{\rm cut}/M_\star$, while the lower axis gives the corresponding
    cumulative mass removed from the surface,
    $\Delta M_{\rm strip}=M_\star-M_{\rm cut}$.
    For every mass cut, the elemental masses are integrated over all
    material exterior to $M_{\rm cut}$ and converted to logarithmic
    elemental number ratios.
    Moving from right to left therefore adds progressively deeper
    stellar layers to the cumulative stripped material.
    Aqua and lavender background regions denote convective and radiative layers,
    respectively, in the terminal structure.
    }
    \label{fig:cumulative_stripping}
\end{figure}

The mass-loss physics of primordial stars in this extreme mass regime remains uncertain, and additional processes could remove substantially more material than the baseline wind prescription considered above. Figure~\ref{fig:cumulative_stripping} therefore poses a conditional question: what chemical composition would become accessible if the terminal $1000\,M_\odot$ model were progressively stripped to greater depths? This is a post-processing accessibility experiment rather than an evolutionary calculation with enhanced mass loss. Increasing the wind prescription itself would alter the stellar mass, rotation, mixing, and subsequent structure and would therefore require the model to be evolved again self consistently. Here we instead integrate the composition of the existing final stellar profile from the surface inward.

The two horizontal axes describe the same stripping depth in complementary ways. The upper axis gives the remaining mass coordinate $M_{\rm cut}/M_\star$, with unity corresponding to the surface and zero to the centre. The lower axis gives the cumulative mass removed,
$\Delta M_{\rm strip}=M_\star-M_{\rm cut}$, which therefore increases from right to left. At each mass cut the elemental mass included in the hypothetical ejecta is
\begin{equation}
M_i^{\rm strip}(M_{\rm cut})
 = \int_{M_{\rm cut}}^{M_\star} X_i(m)\,{\rm d}m
\label{eq:stripped_mass}
\end{equation}
and the vertical axis shows the logarithmic number ratio obtained from all material exterior to that cut. The curves therefore describe cumulative ejecta rather than the local composition at $M_{\rm cut}$.

The main result is how deeply the star must be stripped before this cumulative composition changes. At $M_{\rm cut}/M_\star=0.20$, corresponding to $M_{\rm cut}=195.9\,M_\odot$ and $\Delta M_{\rm strip}=783.9\,M_\odot$, the outer material still gives $\log(N/C)=1.52$, $\log(N/O)=0.65$, and $\log(C/O)=-0.87$. The NeNa and MgAl signatures also remain weak, with $\log(Ne/O)=-3.17$, $\log(Na/O)=-4.88$, and $\log(Al/Mg)=-1.95$. The chemical break begins at $M_{\rm cut}\simeq193.2\,M_\odot$, or $M_{\rm cut}/M_\star\simeq0.197$, after approximately $786.6\,M_\odot$ has been removed. This is the same mass coordinate as the outer O-rich processed-core boundary identified in Figure~\ref{fig:terminal_structure}. It also lies immediately outside the convectively active terminal interior, whose outer boundary occurs near $192.5\,M_\odot$ in the present structure. The flat ratios across the outer $\simeq80\%$ of the stellar mass show that progressively removing the envelope adds material with a very similar composition. This reflects the chemical structure established during the long H and He burning phases together with the subsequent freeze-out of large scale mixing, rather than simply the convective or radiative character of the terminal envelope. The sharp change at $M_{\rm cut}/M_\star\simeq0.2$ occurs only once the stripping reaches the chemically processed interior, where the advanced burning products become incorporated into the cumulative ejecta.

The composition changes abruptly once this interface is crossed. Moving the cut only to $M_{\rm cut}/M_\star\simeq0.19$ raises $\log(Ne/O)$ from about $-3.17$ to $-1.65$ and $\log(Mg/O)$ from $-5.04$ to $-1.93$. The Si signature changes even more strongly, with $\log(Si/O)$ increasing from about $-8.04$ outside the interface to $-2.75$. These changes reflect the addition of Ne, Mg, and Si rich material from the processed interior. In contrast, $\log(Al/Mg)$ falls sharply because Mg increases without a corresponding buildup of Al, while $Na/O$ decreases as increasingly O rich material is included. The transition therefore separates the CNO dominated chemical signature of the envelope from the products of the advanced burning layers.

The Fe group remains confined much deeper. Even after stripping $881.2\,M_\odot$ and leaving only $98.6\,M_\odot$, the cumulative ratio is still only $\log({\rm Fe\,group}/Si)=-4.44$. It rises above $-4$ only once the mass cut falls below about $31.5\,M_\odot$, and reaches $-1.96$ at $M_{\rm cut}=19.3\,M_\odot$. Removing essentially the entire star gives $\log({\rm Fe\,group}/Si)=-0.47$ within the reduced advanced-burning network. Thus, exposing the O, Mg, and Si rich processed region already requires removal of about $80$\% of the terminal star, while appreciable Fe-group material requires stripping to only a few percent of the original remaining mass. The advanced nucleosynthetic products seen in Figure~\ref{fig:elemental_inventory} are therefore not merely absent from the modeled wind, but are physically buried beneath several hundred solar masses of less processed material.

\section{Discussion}
\label{sec:discussion}

\subsection{Angular momentum freeze-in and collapse circularization}
\label{sec:discussion_rotation}

Whether rapid central rotation influences the collapse depends on the specific angular momentum retained by the contracting material rather than on the central angular velocity alone. The rotational evolution shown in Figure~\ref{fig_rotation} demonstrates that the large increase in central angular velocity during advanced burning does not correspond to a comparable increase in the angular momentum of the core. Between core He exhaustion and the common central Si exhaustion profile, the central angular velocity increases by a factor of about $340$, while the total stellar angular momentum decreases by $24.6$\%. Over the same interval the specific angular momentum of fixed Lagrangian shells at enclosed mass coordinates $M_r=3$ and $5\,M_\odot$ changes by only $-0.42$\% and $-0.31$\%, respectively, with a median change of $-0.29$\% across $M_r=0.5$--$5\,M_\odot$. The strong central spin up therefore results primarily from contraction at nearly conserved local $j$ rather than from angular momentum being transported into the inner core. Angular momentum is still redistributed elsewhere in the star and removed through surface mass loss, but these processes do not deliver appreciable additional $j$ to the inner few solar masses during the advanced stages.

The same distinction is apparent when the rotational support of the late core is quantified. At the advanced profile considered here, $\max(\Omega/\Omega_{\rm K})=0.0122$, the maximum ratio of centrifugal to gravitational acceleration is only $1.48\times10^{-4}$, and the global rotational energy ratio is $T_{\rm rot}/|W|=5.88\times10^{-6}$. The rapidly increasing $\Omega_{\rm c}$ seen in Figure~\ref{fig_rotation} therefore reflects the shrinking moment arm of the contracting core while the star remains far from centrifugal support.

\begin{figure}[t]
    \centering
    \includegraphics[width=\columnwidth]{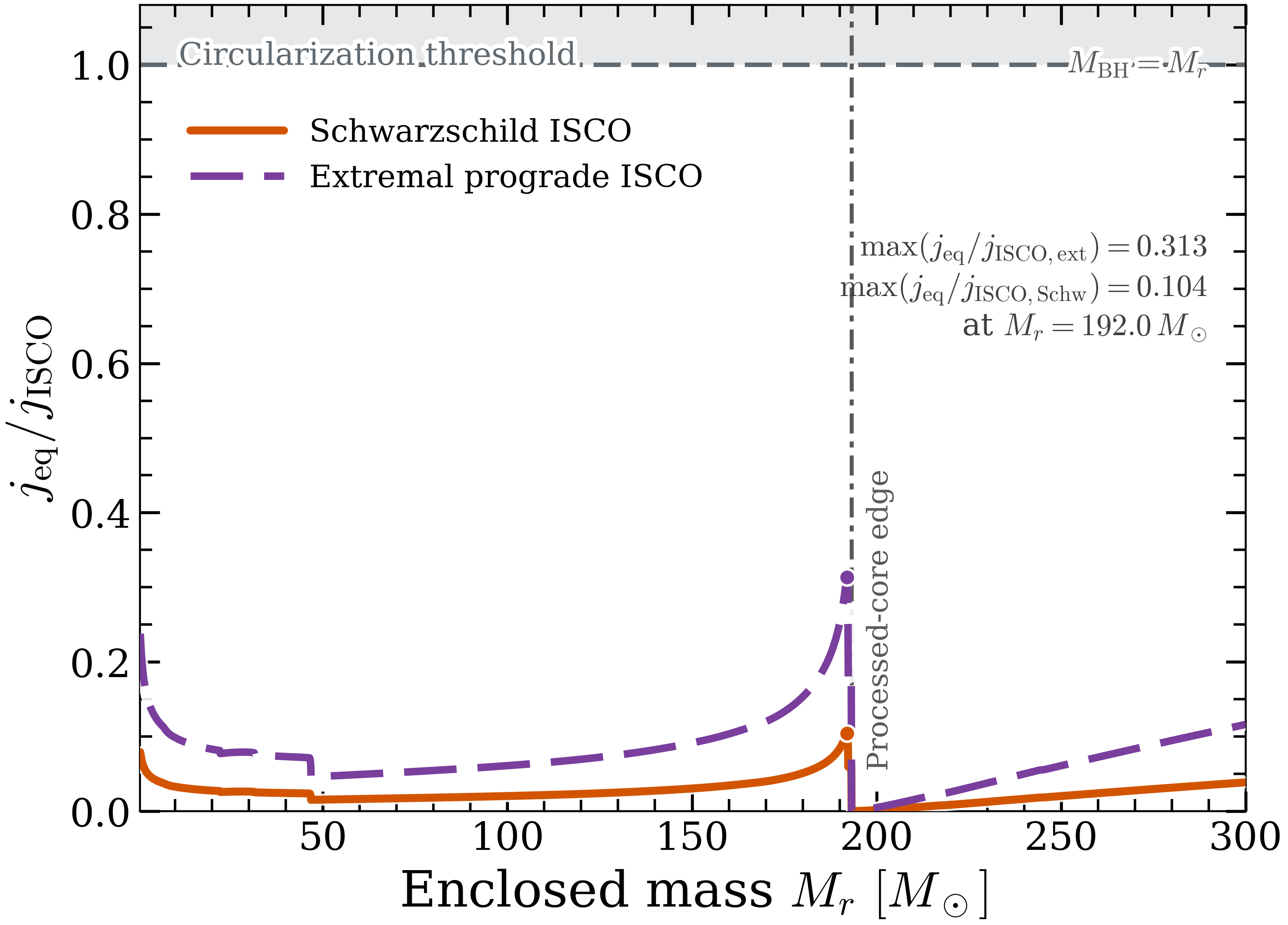}

    \caption{
    Terminal circularization test for the $1000\,M_\odot$ model with
    $v/v_{\rm crit}=0.1$ at the common central Si exhaustion profile.
    The orange solid and purple dashed curves show the equatorial specific
    angular momentum divided by the Schwarzschild and extremal prograde ISCO
    requirements, respectively, assuming $M_{\rm BH}=M_r$.
    The horizontal dashed line marks $j_{\rm eq}/j_{\rm ISCO}=1$.
    The vertical dash-dotted line marks the processed-core boundary, and the
    circles mark the maximum value of each ratio.
    }
    \label{fig:terminal_isco_ratio}
\end{figure}

Figure~\ref{fig:terminal_isco_ratio} tests the more direct collapse requirement by comparing the equatorial specific angular momentum, $j_{\rm eq}=r^2\Omega$, with that required at the prograde innermost stable circular orbit. For a shell initially located at $M_r$, we adopt the inside out approximation $M_{\rm BH}=M_r$, in which the enclosed material has already joined the black hole when that shell accretes. The corresponding limiting values are
\begin{equation}
\begin{array}{rcl}
j_{\mathrm{ISCO,Schw}} &=& \sqrt{12}\,\frac{G M_r}{c}, \\[2pt]
j_{\mathrm{ISCO,ext}}  &=& \frac{2}{\sqrt{3}}\,\frac{G M_r}{c},
\end{array}
\end{equation}
for Schwarzschild and extremal prograde Kerr black holes, respectively.

Following Figure~\ref{fig:terminal_isco_ratio} across the inner $300\,M_\odot$ shows that neither curve reaches the circularization threshold. The largest ratios are $j_{\rm eq}/j_{\rm ISCO,Schw}=0.104$ and $j_{\rm eq}/j_{\rm ISCO,ext}=0.313$, both at $M_r=191.95\,M_\odot$, immediately inside the processed core boundary. The location of this maximum is notable because it lies close to the strong structural and rotational transition identified independently in Figs.~\ref{fig:terminal_structure} and \ref{fig_rotation}. Even there, however, the available specific angular momentum remains below the ISCO requirement by factors of approximately $9.6$ and $3.2$ for the Schwarzschild and extremal limits\footnote{We note that the angular momentum state at the ISCO is also important for possible collapsar-pathways for (ultra-)long gamma-ray bursts (GRBs) originating from massive Pop~III progenitors \citep[e.g.,][]{Bromm_GRB2006,mats15}.}.

The comparison is deliberately favorable to disk formation because Figure~\ref{fig:terminal_isco_ratio} uses the equatorial value $j_{\rm eq}$, which exceeds the shell averaged value $\bar{j}=(2/3)r^2\Omega$. Under the adopted inside out collapse approximation, the computed progenitor therefore contains insufficient specific angular momentum for any shell within the inner $300\,M_\odot$ to circularize at the ISCO.

These estimates should be regarded as progenitor level constraints rather than a calculation of the collapse itself. A dedicated three dimensional relativistic collapse calculation is required to follow black hole growth, hydrodynamic angular momentum redistribution, and any subsequent disk formation self consistently \citep{Uchida2019,Shibata2025}. The present hydrostatic models instead provide the foundational quantities for such calculations, including the retained $j(M_r)$ profile, the degree of rotational support, and the location of the processed core immediately before collapse.

\subsection{CNO self enrichment and radiative mass loss}
\label{sec:cno_mass_loss}

The surface enrichment shown in Figure~\ref{fig:elemental_inventory} raises the possibility that radiative mass loss becomes more important once primary CNO reaches the initially metal free atmosphere. We define $X_{\rm CNO}=X_{\rm C}+X_{\rm N}+X_{\rm O}$, and calculations of CNO driven winds from first stars show that homogeneous outflows can begin near $X_{\rm CNO}\sim10^{-4}$ for favourable stellar parameters even though the corresponding rates remain weak close to this wind limit \citep{Krticka2009}. Our $1000\,M_\odot$ model first reaches $X_{\rm CNO}=1.32\times10^{-4}$ with $2.21\times10^{5}$ yr remaining before the terminal model, when the enrichment is strongly dominated by primary N with $X_{\rm C}=1.40\times10^{-5}$, $X_{\rm N}=1.15\times10^{-4}$, and $X_{\rm O}=2.60\times10^{-6}$. The surface composition changes much more strongly only near the end of He burning, reaching $X_{\rm CNO}=1.04\times10^{-2}$ with $4.69\times10^{3}$ yr remaining and $2.65\times10^{-2}$ at core He exhaustion with only $3.16\times10^{3}$ yr left.

The very different remaining lifetimes make the onset of modest CNO enrichment potentially more important for the integrated radiative mass loss than the much larger abundances reached later. Table~\ref{tab:cno_wind_sensitivity} therefore considers constant rates between $10^{-6}$ and $10^{-3}\,M_\odot\,{\rm yr}^{-1}$ as a sensitivity range rather than as predictions for the present atmosphere. Beginning from the first CNO enriched state, rates of $10^{-6}$, $10^{-5}$, and $10^{-4}\,M_\odot\,{\rm yr}^{-1}$ would remove $0.22$, $2.21$, and $22.1\,M_\odot$, while a sustained rate of $10^{-3}\,M_\odot\,{\rm yr}^{-1}$ would remove $221\,M_\odot$ and necessarily alter the subsequent evolution. By contrast, if these rates operate only after $X_{\rm CNO}$ exceeds $10^{-2}$ then even $10^{-4}\,M_\odot\,{\rm yr}^{-1}$ removes only $0.47\,M_\odot$, while $10^{-3}\,M_\odot\,{\rm yr}^{-1}$ removes $4.69\,M_\odot$ before the terminal model.

\begin{table*}[t]
\centering
\caption{
Sensitivity of the integrated radiative mass loss to surface CNO enrichment.
The final four columns give the mass removed if the indicated constant
$\dot{M}$ operates from each evolutionary point to the terminal hydrostatic
model.
The mass loss rates are in $M_\odot\,{\rm yr}^{-1}$ and the resulting
$\Delta M$ values are in $M_\odot$.
Feedback from the additional mass loss on the subsequent stellar evolution
is not included.
}
\label{tab:cno_wind_sensitivity}

\small
\renewcommand{\arraystretch}{1.25}
\setlength{\tabcolsep}{10pt}

\begin{tabular}{lcccccc}
\hline
\hline
Surface state
&
$X_{\rm CNO}$
&
$\Delta t$ [yr]
&
\multicolumn{4}{c}{$\Delta M\ [M_\odot]$}
\\
\cline{4-7}
&
&
&
$\dot{M}=10^{-6}$
&
$\dot{M}=10^{-5}$
&
$\dot{M}=10^{-4}$
&
$\dot{M}=10^{-3}$
\\
\hline

CNO onset
&
$1.32\times10^{-4}$
&
$2.21\times10^{5}$
&
$0.22$
&
$2.21$
&
$22.1$
&
$221$
\\

Strong CNO enrichment
&
$1.04\times10^{-2}$
&
$4.69\times10^{3}$
&
$4.69\times10^{-3}$
&
$4.69\times10^{-2}$
&
$0.469$
&
$4.69$
\\

Core He exhaustion
&
$2.65\times10^{-2}$
&
$3.16\times10^{3}$
&
$3.16\times10^{-3}$
&
$3.16\times10^{-2}$
&
$0.316$
&
$3.16$
\\

\hline
\hline
\end{tabular}
\end{table*}

The appropriate radiative rate remains uncertain because this CNO enriched atmosphere occupies a regime that is not covered by existing wind calculations. During the relevant evolution the model remains cool at $T_{\rm eff}\simeq7000$--$7200$ K while reaching $\Gamma_{\rm edd}\simeq0.8$--$0.9$, whereas calculations tailored to CNO enriched metal poor stars generally consider substantially hotter atmospheres and lower Eddington factors \citep{Muijres2012}. These calculations also show that CNO enrichment cannot simply be treated as an equivalent scaled solar metallicity because the resulting rates are substantially lower than such a rescaling would imply \citep{Muijres2012}. The high value of $\Gamma_{\rm edd}$ nevertheless favours stronger winds because radiative mass loss rises steeply as VMSs approach the Eddington limit, as found in atmosphere calculations and observations of very massive WNh stars \citep{GraefenerHamann2008,Bestenlehner2014,Bestenlehner2020}. Observed stars in R136 already reach rates of a few $10^{-5}\,M_\odot\,{\rm yr}^{-1}$ at lower $\Gamma_{\rm edd}$, while calculations of inflated $100$--$1000\,M_\odot$ stars at $T_{\rm eff}=15\,000$ K can approach or exceed $10^{-3}\,M_\odot\,{\rm yr}^{-1}$ at finite metallicity \citep{Vink2018}.

The large luminosity of our model also shows that rates near the upper end of Table~\ref{tab:cno_wind_sensitivity} are not excluded by the available photon momentum alone. For a momentum driven wind, $\dot{M}v_\infty=\eta L/c$, where $\eta$ is the wind efficiency, and the luminosity of our model, $\log(L/L_\odot)\simeq7.55$, corresponds to $\dot{M}\simeq1.4$--$3.6\times10^{-3}\,M_\odot\,{\rm yr}^{-1}$ for $\eta=1$ and illustrative terminal velocities of $500$--$200\,{\rm km\,s^{-1}}$. VMS calculations show that $\eta$ can exceed unity once the wind becomes optically thick because multiple interactions allow the outflow momentum to exceed the single scattering value \citep{Vink2011,Vink2012}. The photon momentum budget can therefore support rates of order $10^{-3}\,M_\odot\,{\rm yr}^{-1}$, although this does not imply that such a wind is actually launched. The limiting uncertainty is instead whether a cool atmosphere with strong CNO enrichment but negligible Fe group opacity can couple efficiently enough to the radiation field, since its low effective temperature weakens the UV line driving that underlies the available hot star prescriptions.

The long lived CNO enriched phase therefore places a useful constraint on the missing wind physics. If the radiative rate remains near $10^{-6}$--$10^{-5}\,M_\odot\,{\rm yr}^{-1}$ then the additional mass loss is only a small correction, whereas a sustained rate near $10^{-4}\,M_\odot\,{\rm yr}^{-1}$ would already remove a mass comparable to the surface mass loss in the present calculations. Rates approaching $10^{-3}\,M_\odot\,{\rm yr}^{-1}$ are permitted by the radiative momentum budget but would remove hundreds of solar masses if maintained through the full enriched phase, so such a wind would require the stellar evolution and mass loss to be solved self consistently. The key result is therefore not a unique radiative mass loss rate, but that CNO self enrichment occurs sufficiently early that any sustained wind above approximately $10^{-4}\,M_\odot\,{\rm yr}^{-1}$ would qualitatively change the evolution of these stars.

\subsection{Chemical-mixing freeze-out during advanced burning}
\label{sec:mixing_freezeout}

\begin{figure}[t]
\centering
\includegraphics[width=\columnwidth]{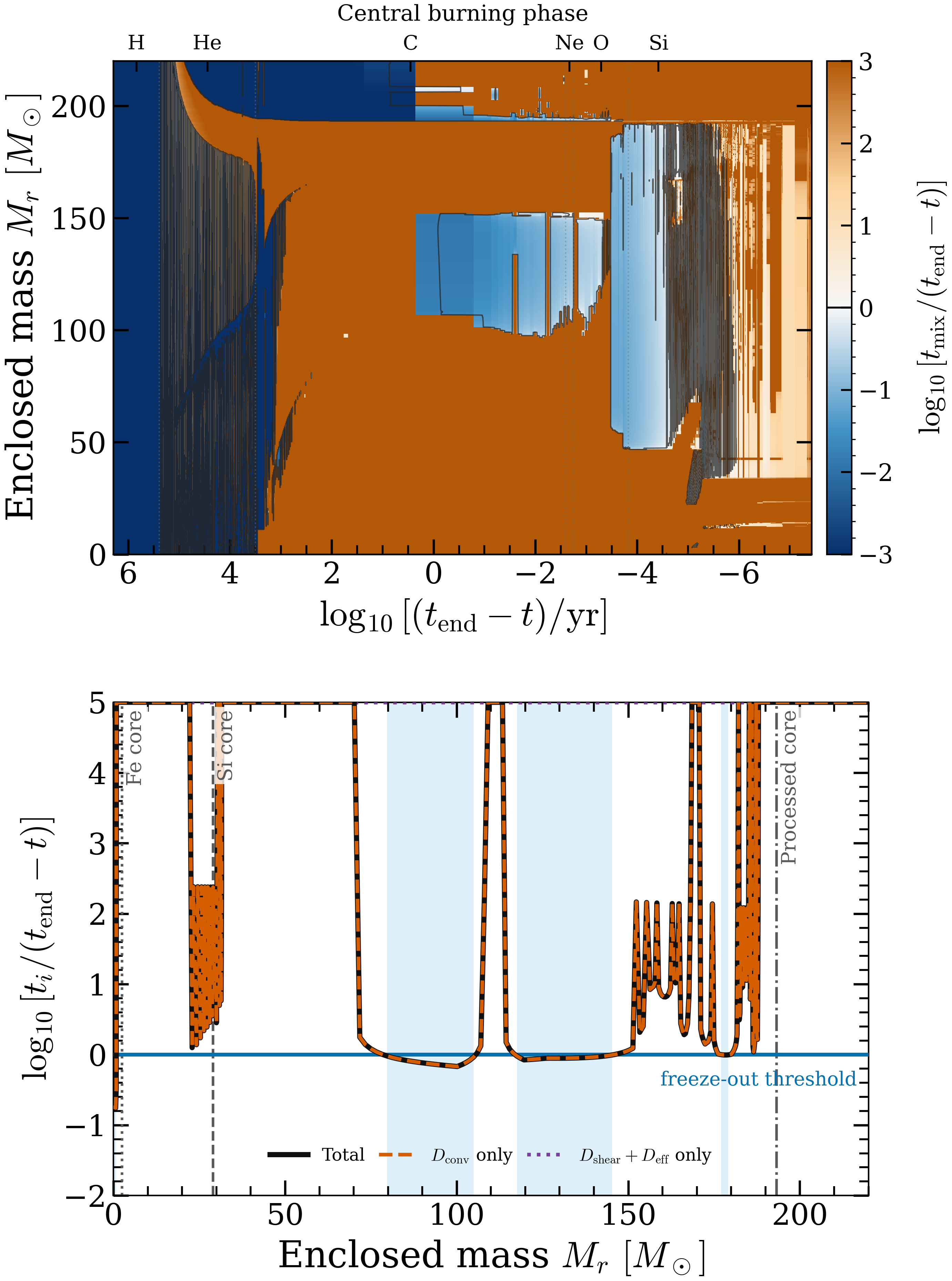}
\caption{
Chemical mixing timescale evolution for the rotating
$1000\,M_\odot$ model.
The upper panel shows
$\log_{10}[t_{\rm mix}/(t_{\rm end}-t)]$ as a function of enclosed mass
and time remaining.
Purple regions have negative values and green regions have positive values.
The black contour marks
$\log_{10}[t_{\rm mix}/(t_{\rm end}-t)]=0$. The lower panel shows $\log_{10}[t_i/(t_{\rm end}-t)]$ at the common central Si exhaustion profile, where $t_i=H_P^2/D_i$ is evaluated for the total coefficient $D_{\rm mix}$, convection alone $D_{\rm conv}$, and rotational transport $D_{\rm shear}+D_{\rm eff}$. The horizontal blue line marks zero and the pale blue regions mark
intervals where the total ratio is negative.
The gray vertical lines mark the Fe core, Si core, and processed core
boundaries.
}
\label{fig:mixing_freezeout}
\end{figure}

The increasingly layered structure produced during advanced burning develops while the remaining evolutionary time contracts from years to minutes. Whether these layers can continue to exchange material therefore depends on the competition between the local transport time and the time available before the final hydrostatic model. Figure~\ref{fig:mixing_freezeout} quantifies this competition for the rotating $1000\,M_\odot$ model. We define
\begin{equation}
D_{\rm mix}=D_{\rm conv}+D_{\rm shear}+D_{\rm eff},
\qquad
t_{\rm mix}=\frac{H_P^2}{D_{\rm mix}} .
\label{eq:mixing_freezeout}
\end{equation}
Here $D_{\rm conv}$ is the convective diffusion coefficient, $D_{\rm shear}$ is the vertical shear diffusion coefficient, $D_{\rm eff}$ is the effective chemical diffusion coefficient associated with meridional circulation, and $H_P$ is the local pressure scale height. The resulting $t_{\rm mix}$ therefore represents the characteristic time to mix material across one pressure scale height. The upper panel shows $\log_{10}[t_{\rm mix}/(t_{\rm end}-t)]$ throughout the available sequence of stored structures. Purple regions have $t_{\rm mix}<t_{\rm end}-t$ and can exchange material across approximately one pressure scale height within the remaining time. Green regions have $t_{\rm mix}>t_{\rm end}-t$ and are locally frozen according to this criterion. The black zero contour marks equality between these timescales and should not be interpreted as a convective boundary.

Figure~\ref{fig:mixing_freezeout} shows that broad connected mixing regions present at earlier times become progressively fragmented as the endpoint is approached. At the common central Si exhaustion profile, only four disconnected intervals satisfy $t_{\rm mix}<t_{\rm end}-t$. These extend from $0$ to $0.81,M_\odot$, from $79.76$ to $104.93,M_\odot$, from $117.55$ to $145.31,M_\odot$, and from $177.07$ to $179.19,M_\odot$. Most of the inner processed structure is therefore unable to redistribute its composition on the remaining evolutionary timescale. At the Si core edge at $M_r=29.03 \, M_\odot$, $t_{\rm mix}=0.857$~d and exceeds the remaining $310.7$~s by a factor of $238$.

The lower panel compares the same dimensionless timescale ratio for the different transport channels at the common central Si exhaustion profile. For each coefficient $D_i$ we define $t_i=H_P^2/D_i$ and plot $\log_{10}[t_i/(t_{\rm end}-t)]$, using $D_{\rm mix}$ for the total transport, $D_{\rm conv}$ for convection, and $D_{\rm shear}+D_{\rm eff}$ for rotational transport. Within the surviving mixing regions, the total timescale closely follows the value obtained from $D_{\rm conv}$. The timescale obtained from $D_{\rm shear}+D_{\rm eff}$ remains above the mixing threshold over these regions. The surviving chemical coupling is therefore maintained primarily by convection, while rotational transport is too slow to reconnect the intervening layers during the final burning stages.

The topology continues to evolve during the final minutes. With $212.9$~s remaining, only three intervals survive from $3.48$ to $4.69 \, M_\odot$, from $26.07$ to $28.45,M_\odot$, and from $34.34$ to $35.34 \, M_\odot$. Their combined width is only $4.60 \, M_\odot$. No sampled shell satisfies the criterion in the profile with $51.9$~s remaining. Chemical mixing therefore freezes through the disappearance and fragmentation of individual mixing regions rather than through one boundary moving smoothly through the star.

This behavior explains why rapid central processing during the final burning stages produces comparatively little change in the global chemical inventory. The large O and N reservoirs established during the earlier evolution remain largely preserved, while the newly synthesized Si and Fe group material remains concentrated in the inner stellar layers. Advanced nucleosynthesis and large scale chemical redistribution therefore become increasingly decoupled as the star approaches its final hydrostatic structure. The surface composition is consequently largely set by the end of He burning, because the remaining advanced burning lifetime is too short for large scale transport to communicate newly synthesized material from the compact interior to the envelope.

\subsection{Comparison with previous work}
\label{sec:comparison_previous}

The closest calculations in initial mass are those of \citet{Ohkubo2006}, who evolved non-rotating, zero-metallicity $500$ and $1000\,M_\odot$ stars through iron-core collapse. Their $1000\,M_\odot$ model reaches $\rho_{\rm c}\sim10^{10}\,{\rm g\,cm^{-3}}$, compared with $\log(\rho_{\rm c}/{\rm g\,cm^{-3}})=8.691$ in our final hydrostatic structure. They also obtain an iron core of about $250\,M_\odot$, although this is not directly comparable with either our $193.2\,M_\odot$ O-rich processed region or the $17.0\,M_\odot$ Fe-group inventory because the definitions and evolutionary stages differ. The comparison nevertheless shows a consistent physical progression. Our models resolve the formation of the advanced-burning core, while the substantially denser structures of Ohkubo et al. follow its subsequent contraction toward collapse.

The evolutionary endpoint is also sensitive to how pair softening is defined. \citet{volp23} terminate their primordial VMS models when a mass-weighted $\langle\Gamma_1\rangle$ falls below $4/3+0.01$. Their $1000\,M_\odot$ model therefore stops much earlier in the advanced evolution, whereas our calculation reaches central Si exhaustion with $\langle\Gamma_1\rangle_{P\,dV}=1.34765$ and $\Gamma_{1,\min}=1.34082$, while no shell reaches $\Gamma_1<4/3$. The difference should not be attributed to rotation alone. Their $1000\,M_\odot$ model is nonrotating, whereas ours includes rotation, but the calculations also differ in mass loss and the resulting stellar structure. In addition, their adopted criterion identifies significant pair softening before local dynamical instability, whereas we continue the hydrostatic evolution and test explicitly whether any layer crosses the $\Gamma_1=4/3$ boundary. Rotation may therefore contribute indirectly through its effect on the stellar structure, but the present comparison does not isolate it as the cause of the different $\Gamma_1$ evolution.

The early nuclear evolution is considerably less sensitive than the envelope evolution. The rotating $500\,M_\odot$ models of \citet{Yusof2013} have H-burning lifetimes of $2.132$ and $2.077$ Myr at solar and LMC metallicity, respectively, compared with $2.1225$ Myr here. The agreement is within $2.2\%$, despite radically different retained masses. Their models are reduced to $24.6$ and $101.7\,M_\odot$ by H exhaustion, whereas our primordial model still contains $499.8\,M_\odot$ at the final hydrostatic stage. The zero-metallicity rotating models of \citet{Yoon2012} likewise give a similar H-burning lifetime of $2.07$ Myr but retain only $413.3\,M_\odot$. The nuclear clock is therefore largely set by the enormous mixed core and luminosity of a VMS, while metallicity-dependent winds, rotational mass loss, and angular momentum transport determine whether the envelope is retained.

The strongest structural sensitivity appears after He exhaustion. Between our $900$ and $1000\,M_\odot$ models, $X_{\rm c}(^{12}{\rm C})$ decreases from $0.100$ to $0.063$ and the central density from $\log\rho_{\rm c}=2.933$ to $2.768$. Before accounting for the temperature dependence of the reaction rate, the factor $\rho X_{\rm C}^{2}$ entering
\[
\epsilon_{\rm C}\propto\rho X_{\rm C}^{2}\langle\sigma v\rangle
\]
is therefore only $0.27$ as large in the $1000\,M_\odot$ model. Its slightly higher central temperature partly compensates for this through the strongly temperature-sensitive carbon fusion rate, but the balance between contraction and C burning is substantially altered. By Si exhaustion the O-rich processed boundary consequently moves from $87.7$ to $193.2\,M_\odot$, a factor of $2.2$ change for only an $11\%$ increase in initial mass. The sensitivity of later C and O burning to the carbon abundance inherited from He burning is well established in lower-mass presupernova calculations \citep{Sukhbold2014,Patton2020}. Our models show that the same physics can produce structural differences exceeding $100\,M_\odot$ in extremely massive stars.

Rotation produces an equally important distinction between central spin-up and retained collapse angular momentum. As in classical rotating presupernova calculations \citep{Heger2000}, contraction drives a rapid increase in $\Omega$ once the nuclear timescale becomes shorter than the transport timescale. In our $1000\,M_\odot$ model, $\Omega_{\rm c}$ increases by a factor of about $340$ between He and Si exhaustion, while the specific angular momentum within the inner $5\,M_\odot$ changes by less than 1\%. The core therefore spins up almost entirely through contraction rather than inward angular momentum transport. It nevertheless remains far from rotational support, with $\max(\Omega/\Omega_{\rm K})=0.0122$ and a maximum $j_{\rm eq}/j_{\rm ISCO}=0.313$ even for an extremal prograde black hole. This differs from the disk-compatible primordial models of \citet{Volpato2024}, which begin with substantially larger retained specific angular momentum. The relevant quantity for collapse is therefore not the large central angular velocity itself, but whether the frozen-in $j$ exceeds the circularization requirement of the growing black hole.

Finally, the chemical evolution separates nucleosynthesis from chemical feedback. Primary-N production in rotating primordial stars has also been found in previous calculations \citep{Nandal2024,Tsiatsiou2024}. Our $1000\,M_\odot$ model produces $62.1\,M_\odot$ of N by He exhaustion, but loses only $20.2\,M_\odot$ through the modeled wind. The advanced-burning products are even less accessible: approximately $786.6\,M_\odot$ must be removed before stripping reaches the $193.2\,M_\odot$ processed-core boundary. This follows directly from the hierarchy of evolutionary and transport timescales. Primary N is produced while convection and rotational mixing operate for $\sim2.4\times10^{5}$ yr during He burning, whereas the later Ne, O, and Si-burning stages proceed over days to hours and become chemically frozen. These stars can therefore synthesize substantial heavy-element reservoirs without ejecting them, making the depth of mass removal as important as the nucleosynthesis itself for their chemical contribution.

\section{Summary and Conclusions}
\label{sec:conclusions}

We have investigated for the first time the advanced evolution of rotating Pop III stars with initial masses from $500$ to $1000\,M_\odot$ using the extended GENEC stellar evolution framework. Our standard sequence assumes $v/v_{\rm crit}=0.1$ and is followed through the advanced nuclear burning stages, while an additional $1000\,M_\odot$ model with $v/v_{\rm crit}=0.4$ is evolved through core He exhaustion. We have used these calculations to determine how rotation and nuclear burning shape the final internal structure, angular momentum distribution and chemical inventory of extremely massive primordial stars. We find that

\begin{itemize}

\item
The stars begin their evolution with exceptionally large convective cores that enclose about $93$\% to $95$\% of the initial stellar mass. The $800$, $900$ and $1000\,M_\odot$ models begin their strong redward evolution before central H exhaustion, whereas the less massive models remain hot until He burning has begun.

\item
The evolutionary timescale falls from millions of years during H burning to hours during Si burning. In the $1000\,M_\odot$ model the C burning interval lasts $3.16$ kyr while Ne and O burning last only $0.282$ and $0.593$ days, followed by a Si burning interval of $1.21$ hr. The resulting advanced core is extremely compact even though the stellar envelope remains several thousand solar radii across, with the inner $100\,M_\odot$ contained inside only $0.088\,R_\odot$.

\item
The terminal structure does not vary smoothly with initial mass. Between $900$ and $1000\,M_\odot$ the residual central carbon abundance after He burning decreases from $0.100$ to $0.063$, while the O-rich processed boundary at common Si exhaustion moves from $87.7$ to $193.2\,M_\odot$. The processed region therefore increases by a factor of $2.2$ for only an $11$\% increase in initial mass, showing that the conditions inherited from He burning can dominate the later C and O burning structure. 

\item
The $1000\,M_\odot$ model approaches the thermodynamic region associated with pair softening without developing a locally pair unstable core during the calculated hydrostatic evolution. At central Si exhaustion the minimum value is $\Gamma_{1,\min}=1.34082$ and the pressure volume weighted value is $\langle\Gamma_1\rangle_{P\,dV}=1.34765$, while no shell reaches $\Gamma_1<4/3$. The central trajectory alone therefore does not determine whether the stellar structure has entered dynamical pair instability.

\item
The rapid rotation of the late core is produced primarily by contraction rather than by inward angular momentum transport. Between He and Si exhaustion the central angular velocity increases by a factor of $392$, while the specific angular momentum in the inner few solar masses changes by less than 1\% and the total stellar angular momentum decreases by $24.6$\%. The final core nevertheless remains far from rotational support with $\max(\Omega/\Omega_{\rm K})=0.0122$, while the largest value of $j_{\rm eq}/j_{\rm ISCO}$ is only $0.313$ even for the extremal prograde limit. Under our inside out collapse assumption the computed core therefore does not contain enough specific angular momentum to circularize into a disk. 

\item
Rotation also leaves a strong chemical imprint before the advanced burning stages become dynamically rapid. The $1000\,M_\odot$ model builds $62.1\,M_\odot$ of primary N by He exhaustion and this reservoir changes by less than 1\% thereafter, while the terminal structure still contains about $125\,M_\odot$ of O together with $52.5\,M_\odot$ in the Si group and $17.0\,M_\odot$ in the Fe group. These quantities describe the hydrostatic elemental reservoirs rather than ejecta yields, since most of the advanced material remains confined to the compact interior. 

\item
The separation between production and ejection becomes especially clear when the transport timescales are considered. At the edge of the Si core the mixing time exceeds the remaining evolutionary time by a factor of $238$, while the surviving mixed regions become progressively fragmented and disappear from the sampled structure during the final minute. The same separation is seen chemically because about $786.6\,M_\odot$ must be removed from the terminal $1000\,M_\odot$ model before the stripping reaches the O rich processed boundary, while appreciable Fe group material is buried considerably deeper. 

\end{itemize}

These results reveal an increasingly strong separation between the core and envelope as extremely massive primordial stars approach the end of hydrostatic evolution. The long H and He burning phases establish the global chemical and angular momentum structure, while the rapidly shortening advanced stages process an increasingly compact central region without having sufficient time to reorganize the rest of the star. The resulting progenitors can therefore contain large reservoirs of newly synthesized material and rapidly rotating central layers while still retaining an extended envelope and insufficient specific angular momentum for centrifugal support during direct collapse.

Several extensions will be important for determining how general these conclusions remain. Future calculations should extend the grid toward higher metallicity and larger initial rotation rates, while the inclusion of magnetic angular momentum transport through the Taylor-Spruit dynamo will test how strongly the retained core angular momentum changes. The final hydrostatic structures should also be evolved with a dedicated collapse code so that black hole formation and possible disk formation can be determined dynamically, rather than inferred from the progenitor structure alone.

\begin{acknowledgments}
DN was supported by the Swiss National Science Fund (SNSF) Postdoctoral Fellowship, grant number: P500-2235464. Devesh Nandal would like to thank Mr. John Michael Godier for a stimulating discussion during the Event Horizon Show, which eventually inspired this work. 
AACS is supported by the Deutsche Forschungsgemeinschaft (DFG, German Research Foundation) in the form of an Emmy Noether Research Group – Project-ID 445674056 (PI Sander). AACS further acknowledges support from the Federal Ministry of Education and Research (BMBF) and the Baden-W{\"u}rttemberg Ministry of Science as part of the Excellence Strategy of the German Federal and State Governments. This project was co-funded by the European Union (Project 101183150 - OCEANS). 
\end{acknowledgments}

\bibliography{sample7}{}
\bibliographystyle{aasjournalv7}



\end{document}